\documentclass[aps,prl,twocolumn,superscriptaddress,groupedaddress, floatfix]{revtex4}
\usepackage[utf8]{inputenc}
\usepackage{amsmath}
\usepackage{graphicx}
\usepackage{float}
\usepackage{color}
\usepackage[paperwidth=210mm,paperheight=297mm,centering,hmargin=1.5cm,vmargin=2.0cm]{geometry}

\usepackage{natbib}
\setcitestyle{nature}

\graphicspath{{Figures/}}

\newcommand{\fa}{f_{\downarrow}}
\newcommand{\fb}{f_{\uparrow}}

\newcommand{\fr}{f_{\mathrm{r}}}

\newcommand{\Tp}{T_\mathrm{parity}}
\newcommand{\Ts}{T_\mathrm{s}}
\newcommand{\ket}[1]{| #1 \rangle}
\newcommand{\bra}[1]{\langle #1 |}

\newcommand{\nm}{\mathrm{nm}}
\newcommand{\mm}{\mathrm{mm}}

\newcommand{\um}{\mathrm{\mu}\mathrm{m}}

\newcommand{\us}{\mathrm{\mu}\mathrm{s}}

\newcommand{\ns}{\mathrm{ns}}

\newcommand{\g}{\ket{g}}
\newcommand{\oa}{\ket{\! \downarrow_q \,}}
\newcommand{\ob}{\ket{\! \uparrow_q \,}}
\newcommand{\oC}{\ket{\! \uparrow_a \,}}
\newcommand{\od}{\ket{\! \downarrow_a \,}}

\newcommand{\fd}{f_\mathrm{d}}
\newcommand{\Vg}{V_\mathrm{g,c}}
\newcommand{\Vnw}{V_\mathrm{g,p}}

\newcommand{\GHz}{\mathrm{GHz}}
\newcommand{\MHz}{\mathrm{MHz}}

\newcommand{\mV}{\mathrm{mV}}
\newcommand{\meV}{\mathrm{meV}}
\newcommand{\es}{\epsilon_\mathrm{s}}
\newcommand{\TR}{T_{2R}}
\newcommand{\TE}{T_{2E}}

\makeatletter
\renewcommand{\p@subsection}{}
\renewcommand{\p@subsubsection}{}
\makeatother

\begin{document}

\widetext

\title{Coherent manipulation of an Andreev spin qubit}

\author{M.~Hays}
\email{max.hays@yale.edu}
\affiliation{Department of Applied Physics, Yale University, New Haven, CT 06520, USA}
\author{V.~Fatemi}
\email{valla.fatemi@yale.edu}
\affiliation{Department of Applied Physics, Yale University, New Haven, CT 06520, USA}
\author{D.~Bouman}
\affiliation{QuTech and Delft University of Technology, 2600 GA Delft, The Netherlands}
\affiliation{Kavli Institute of Nanoscience, Delft University of Technology, 2600 GA Delft, The Netherlands}
\author{J.~Cerrillo}
\affiliation{Área de F\'isica Aplicada, Universidad Polit\'ecnica de Cartagena, E-30202 Cartagena, Spain}
\affiliation{Departamento de F\'isica Te\'orica de la Materia Condensada C-V, Universidad Aut\'onoma de Madrid, E-28049 Madrid, Spain}
\author{S.~Diamond}
\affiliation{Department of Applied Physics, Yale University, New Haven, CT 06520, USA}
\author{K.~Serniak}
\affiliation{Department of Applied Physics, Yale University, New Haven, CT 06520, USA}
\affiliation{Current Affiliation: MIT Lincoln Laboratory, 244 Wood Street, Lexington, MA 02420, USA}
\author{T.~Connolly}
\affiliation{Department of Applied Physics, Yale University, New Haven, CT 06520, USA}
\author{P.~Krogstrup}
\affiliation{Center for Quantum Devices, Niels Bohr Institute, University of Copenhagen, Universitetsparken 5, 2100 Copenhagen, Denmark}
\author{J.~Nygård}
\affiliation{Center for Quantum Devices, Niels Bohr Institute, University of Copenhagen, Universitetsparken 5, 2100 Copenhagen, Denmark}
\author{A.~Levy~Yeyati}
\affiliation{Departamento de F\'isica Te\'orica de la Materia Condensada C-V, Universidad Aut\'onoma de Madrid, E-28049 Madrid, Spain}
\affiliation{Condensed Matter Physics Center (IFIMAC) and Instituto Nicol\'as Cabrera, Universidad Aut\'onoma de Madrid, E-28049 Madrid, Spain}
\author{A.~Geresdi}
\affiliation{QuTech and Delft University of Technology, 2600 GA Delft, The Netherlands}
\affiliation{Kavli Institute of Nanoscience, Delft University of Technology, 2600 GA Delft, The Netherlands}
\affiliation{Quantum Device Physics Laboratory, Department of Microtechnology and Nanoscience, Chalmers University of Technology, SE 41296 Gothenburg, Sweden}
\author{M.~H.~Devoret}
\email{michel.devoret@yale.edu}
\affiliation{Department of Applied Physics, Yale University, New Haven, CT 06520, USA}


\begin{abstract}
 Two promising architectures for solid-state quantum information processing are electron spins in semiconductor quantum dots and the collective electromagnetic modes of superconducting circuits. 
 In some aspects, these two platforms are dual to one another: superconducting qubits are more easily coupled but are relatively large among quantum devices $(\sim\mm)$, while electrostatically-confined electron spins are spatially compact ($\sim\um$) but more complex to link.
 Here we combine beneficial aspects of both platforms in the Andreev spin qubit: the spin degree of freedom of an electronic quasiparticle trapped in the supercurrent-carrying Andreev levels of a Josephson semiconductor nanowire.
We demonstrate coherent spin manipulation by combining single-shot circuit-QED readout and spin-flipping Raman transitions, finding a spin-flip time $\Ts = 17~\us$ and a spin coherence time $\TE=52~\ns$. 
These results herald a new spin qubit with supercurrent-based circuit-QED integration and further our understanding and control of Andreev levels -- the parent states of Majorana zero modes -- in semiconductor-superconductor heterostructures.
\end{abstract}

\maketitle

\vspace{-10mm}

A weak link between two superconductors hosts discrete, fermionic modes known as Andreev levels~\cite{beenakker1991june, Furusaki1991}.
They govern the physics of the weak link on the microscopic scale, ultimately giving rise to macroscopic phenomena such as the Josephson supercurrent.
Superconducting quantum circuits crucially rely on the nonlinearity of the supercurrent in Josephson tunnel junctions, a manifestation of the ground-state properties of millions of Andreev levels acting in concert~\cite{clarke2004,devoret2013,Roy2016}.
While the vast majority of conduction electrons participate in the nonlinear bosonic oscillations of the superconducting condensate, each Andreev level is itself a fermionic degree of freedom, able to be populated by electronic excitations known as Bogoliubov quasiparticles.

In 2003, it was proposed to store quantum information in the spin state of a quasiparticle trapped in a weak link possessing a spin-orbit coupling~\cite{chtchelkatchev2003andreev, padurariu2010theoretical, reynoso2012spin,park2017andreev}.
It was pointed out that this Andreev spin qubit would carry a state-dependent supercurrent, opening new paths for spin manipulation and measurement that are unavailable to electrostatically-confined spin qubits~\cite{hanson2007spins, childress2013diamond}. 
In particular, such a state-dependent supercurrent could be used to achieve strong coupling with a superconducting microwave resonator, an area of active research in the spin qubit community~\cite{petersson2012circuit, samkharadze2018strong, mi2018coherent, harvey2018coupling, landig2018coherent, cubaynes2019highly, borjans2020resonant}. 
This supercurrent-based coupling has been used in such circuit quantum electrodynamics (cQED) architectures~\cite{blais2004cavity, wallraff2004strong} to detect and manipulate pairs of quasiparticles trapped in Andreev levels~\cite{Janvier15,hays2018direct}. 
However, because the Andreev levels of most weak links are paired into spin-degenerate doublets, quasiparticle spin manipulation has remained out of reach. 

The level structure of an Andreev doublet is determined by the geometric and material properties of the host weak link, as shown in weak links composed of superconductor-proximitized semiconductor nanowires, or ``Josephson nanowires'' for short~\cite{Krogstrup15,chang2015hard}. 
Thanks to recently-achieved atomic-scale perfection of the superconductor-semiconductor interface, it is now possible to observe the Andreev spectra of Josephson nanowires, revealing a rich interplay between electromagnetic field effects, device geometry, and spin-orbit coupling~\cite{van2017microwave, tosi2019spin}.
These properties of Andreev levels in superconductor-semiconductor nanowires have been employed to demonstrate gate-tunable weak links for superconducting qubits~\cite{larsen2015semiconductor,de2015realization}, probe non-abelian Andreev levels known as Majorana zero modes~\cite{fu2008superconducting,Lutchyn2010, Oreg2010, Mourik2012, deng2016majorana}, and, importantly for this experiment, investigate spin-split doublets without a Zeeman field~\cite{tosi2019spin, hays2020continuous}. 

In this letter, we demonstrate the first coherent manipulation of the spin of an individual quasiparticle excitation of a superconductor. 
The quasiparticle is trapped in the Andreev levels of a Josephson nanowire, where it resides predominantly in the two spin states of the lowest-energy Andreev doublet with roughly equal probability. 
First, we initialize this Andreev spin qubit in one of the two spin states by post-selecting on a single-shot cQED spin measurement, which we demonstrated in an earlier work~\cite{hays2020continuous}. 
We then achieve full coherent control of the Andreev spin qubit by driving Raman transitions in a natural $\Lambda$ system formed by the two spin states and an excited state. 
We observe spin lifetimes up to $\Ts = 17~\us$ at the presented gate voltages [see Supplementary Information Fig. S9].
However, the much shorter spin coherence time $T_2=52~\ns$ appears to be limited by a spinful bath. 

\begin{figure}[H]
	\centering
	\includegraphics[width=\columnwidth]{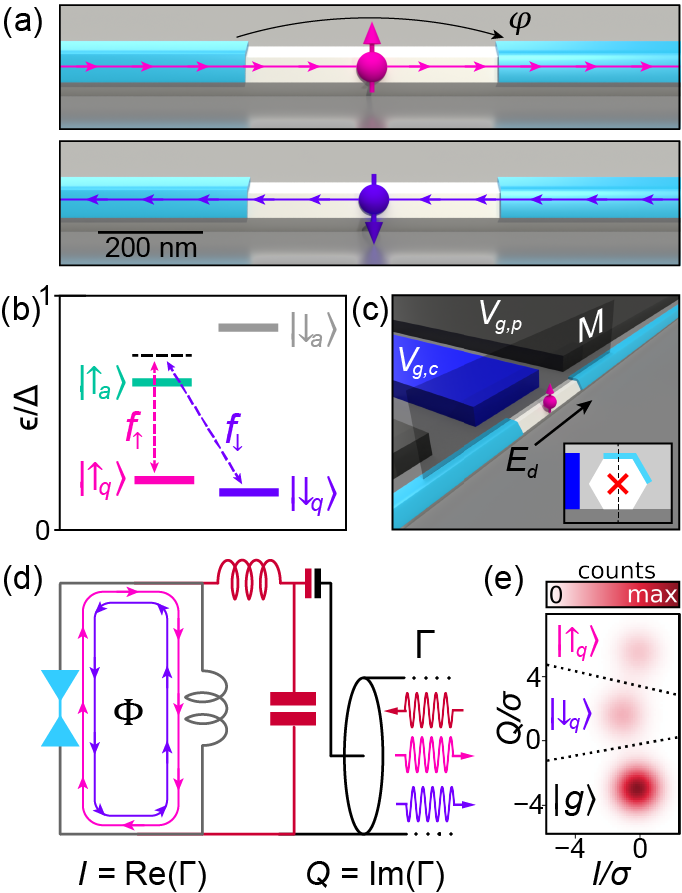} 
	\caption{
    Principle of the Andreev spin qubit.
	(a)
	Illustration of a semiconductor nanowire (white) coated with epitaxial superconductor (light blue). 
	A quasiparticle is trapped in the exposed weak link by the pair potential $\Delta$ of the superconducting leads.
	Due to spin-orbit coupling, if the quasiparticle is in the spin-up state (upper panel) supercurrent flows to the right near zero phase bias $\varphi\ne0$, while in the spin-down state (lower panel) supercurrent flows to the left. 
	Applying nonzero $\varphi$ thus breaks spin degeneracy. 
	(b) 
	Level structure of two Andreev doublets tuned to a $\Lambda$ configuration. 
    Two microwave drives (frequencies $f_\downarrow$ and $f_\uparrow$) are equally detuned from $\oa \leftrightarrow \oC$ and $\ob \leftrightarrow \oC$, inducing a coherent Raman process between the qubit states $\oa$ and $\ob$ via a virtual level (black dashed line).
	(c)
	Both microwave drives induce an rf electric field $E_\mathrm{d}$ between the superconducting leads. 
	But, for a nanowire symmetric across the plane $M$ (i.e. only nanowire + substrate), drive-induced spin-flips would be forbidden. 
	However, as depicted in the inset, here the mirror symmetry is broken by both the partial aluminum shell as well as the cutter (blue, bias $\Vg$) and plunger gates (black, bias $\Vnw$).
	(d) 
	The Josephson nanowire (light blue) is embedded in a superconducting loop (gray), which enables phase bias via an external flux $\varphi \approx 2\pi\Phi/\Phi_0$ as well as inductive coupling to a superconducting microwave resonator (maroon).
    The resonator reflection coefficient $\Gamma = I + i Q$ is probed with a tone near the resonator frequency $\fr = $ 9.188 GHz. 
    (e)
    Repeated $1.9~\us$ measurements of $I$ and $Q$ clustered into three distributions, corresponding to $\oa$, $\ob$ and $\g$ (standard deviation $\sigma$). 
    The system state was assigned based on thresholds indicated by the black dotted lines.
}
\end{figure}

Our realization of the Andreev spin qubit hinges on the interplay between spin-orbit coupling in the semiconductor nanowire and the superconducting phase bias $\varphi$ across the weak link [Fig.~1(a)]~\cite{chtchelkatchev2003andreev, padurariu2010theoretical, reynoso2012spin,park2017andreev,tosi2019spin, hays2020continuous}. 
In a conventional weak link, a trapped quasiparticle is restricted to spin-degenerate Andreev doublets and therefore the spin cannot be coherently manipulated. 
In a Josephson nanowire, however, an inter-subband spin-orbit interaction can cause spin to hybridize with translational degrees of freedom (this hybridized spin is sometimes known as pseudospin, though we will continue to refer to it as ``spin'' for simplicity). 
Due to this interaction between spin and motion, the two spin states of an Andreev doublet carry equal and opposite supercurrent $\pm I_\mathrm{s}/2$ at $\varphi=0$, with $I_\mathrm{s}$ doublet-dependent. 
The doublet degeneracy can thus be lifted with a nonzero phase bias: perturbatively near $\varphi=0$, the spin splitting is given by $\es = I_\mathrm{s} \varphi \,\Phi_0/2\pi $.

Microwave quantum optics techniques are well-suited to achieve quasiparticle spin manipulation, given the frequency selectivity brought about by such a flux-induced spin splitting. 
In this experiment, the two spin states $\oa, \ob$ of one Andreev doublet form the qubit basis [Fig.~1(b)], while a second, higher-energy doublet provides auxiliary states $\oC$, $\od$ critical for both qubit control and measurement~\cite{hays2020continuous}.
To manipulate the Andreev spin qubit, we use the qubit states $\oa,\ob$ in conjunction with $\oC$ as a $\Lambda$ system.
We apply simultaneous microwave drives to both the $\ob \leftrightarrow \oC$ transition (drive frequency $f_\uparrow$) and $\oa \leftrightarrow \oC$ (drive frequency $f_\downarrow$). 
By equally detuning the two drives from their respective transitions, a Raman process is induced such that the $\{\oa, \ob\}$ manifold can be coherently manipulated while $\oC$ remains minimally populated.

The success of the Raman process is contingent on our ability to drive both the spin-conserving transition $\ob \leftrightarrow \oC$ and the spin-flipping transition $\oa \leftrightarrow \oC$.
While spin-orbit hybridization is necessary to enable electric-field induced spin-flips~\cite{nadj2010spin}, in our situation a broken spatial symmetry of the Josephson nanowire is also required (see Supplementary Information for details).
In this experiment, our hexagonal nanowire was made of [001] wurtzite indium arsenide grown by molecular beam epitaxy.  
Such a nanowire lying alone on a substrate would possess a transverse mirror symmetry [Fig. 1(c)]; this property would then be inherited by the levels of the nanowire such that one spin state of each doublet would be mirror-symmetric and the other anti-symmetric.
Since we apply the microwave drive voltage across the weak link, the rf electric field respects the mirror symmetry (it points along the nanowire) and therefore cannot flip spin.

In the device used in this work, the mirror symmetry is broken by both the superconducting leads and the electrostatic gates [Fig. 1(c)], as well as any symmetry-breaking disorder present in the nanowire.  
The superconducting leads consist of $10~\nm$-thick epitaxial aluminum, of which a $500~\nm$ length was removed to form the weak link.
The aluminum only covers two of six nanowire facets, thereby breaking the mirror symmetry of the nanowire-substrate system. 
As the gates are fabricated on one side of the nanowire, they also break the mirror symmetry. 
Both the cutter and plunger gates were used to tune the transparency of the weak link and were biased to $\Vg = -71.9~\mV$ and $\Vnw = 4.0~\mV$ respectively, unless otherwise noted (see Supplementary Information for system tune up). 
With the mirror symmetry broken, the drive may induce spin flips and a Raman process can be used for coherent spin manipulation. 

We detect the state of the Andreev spin qubit by embedding the Josephson nanowire in a cQED architecture [Fig.~1(d)]. 
As we previously demonstrated~\cite{hays2020continuous}, the effect of spin-orbit coupling on both the inter-doublet transition spectrum and supercurrent can be harnessed to achieve a spin-dependent dispersive shift of the frequency of a superconducting microwave resonator. 
Following that work, we detect the quasiparticle spin state by measuring the resonator response to a resonant probe tone [Fig.~1(e)]. 
The complex amplitude $\Gamma=I+iQ$ of the reflected tone clustered into three distributions, corresponding to the two spin states $\oa$, $\ob$ of a trapped quasiparticle as well as the ground state $\g$ of the junction where no quasiparticle was present. 
Throughout this work, we present data in terms of spin state occupation probabilities $P_\uparrow, P_\downarrow$, which we compute based on the thresholds displayed in Fig. 1(e). 

The Andreev spin qubit exists exclusively when a quasiparticle is stochastically trapped in the Josephson nanowire (see Supplementary Information for a quantum jumps trace). 
For the bias conditions presented in this work, we found that a trapped quasiparticle occupied the two spin states of the lowest doublet with roughly equal probability. 
Thus, under any coherent manipulation $\oa \leftrightarrow \ob$ the observed spin state populations would not change.
Throughout this work, we overcame this problem by initializing the quasiparticle in $\ob$ via an initial readout pulse and post-selection (see Supplementary Information for the same measurements with $\oa$ post-selection).
Our single-shot spin readout was thus critical to our observation of coherent population transfer between $\ob$ and $\oa$.

\begin{figure}[h]
	\centering
	\includegraphics[width=\columnwidth]{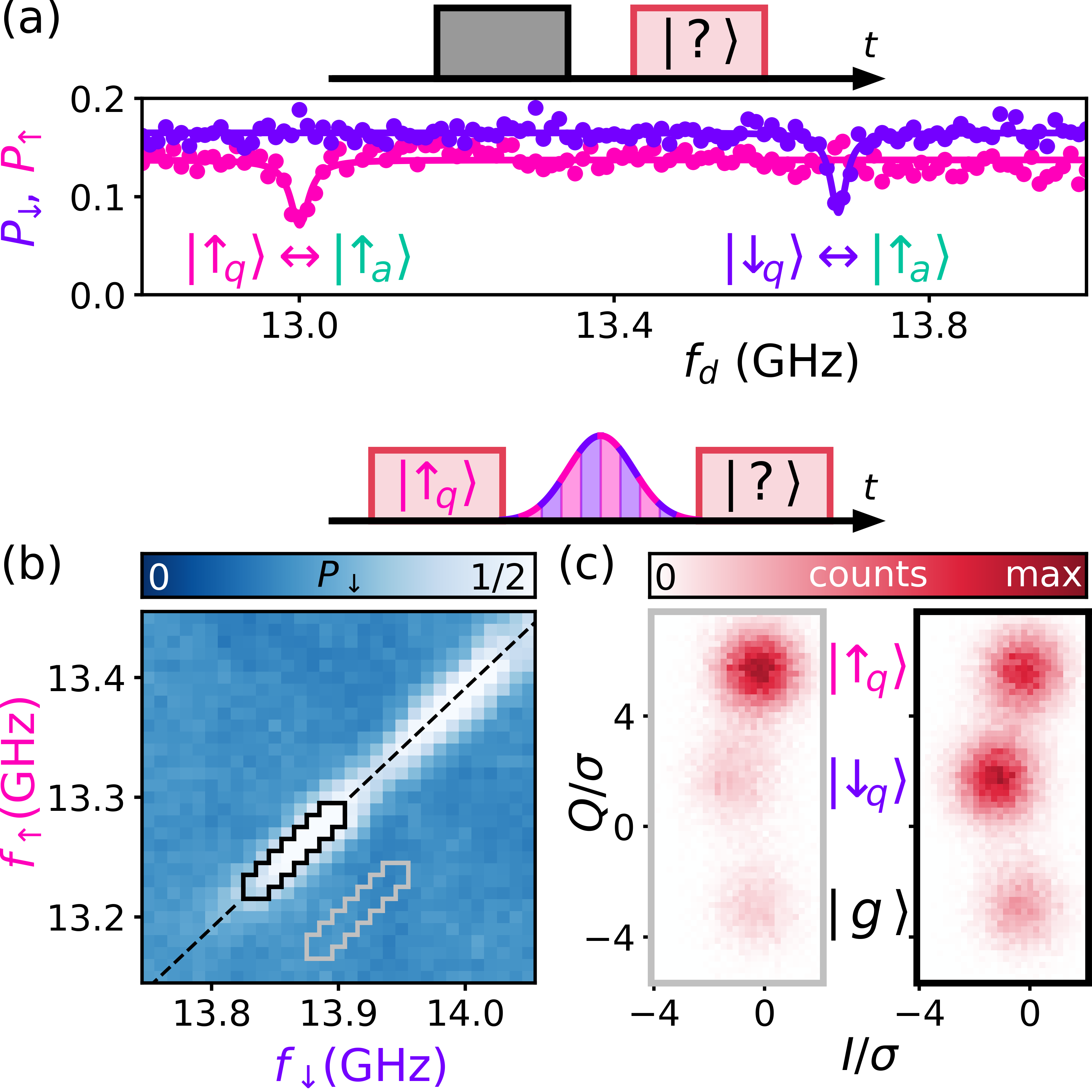} 
	\caption{
    Raman transitions of a trapped quasiparticle. 
    (a)
    In a two-tone measurement, the Josephson nanowire was first driven by a $1~\us$ saturation pulse (gray) of variable carrier frequency $f_d$ before the quasiparticle state was determined with a readout pulse (maroon). 
    A dip is observed in $P_\uparrow$ corresponding to the $\ob \leftrightarrow \oC$ transition and in $P_\downarrow$ corresponding to the $\oa \leftrightarrow \oC$ transition [see Fig. S5 for $\Phi$-dependence].
    (b)
    The quasiparticle was first prepared in $\ob$ via an initial readout pulse and post-selection.
    Simultaneous Gaussian pulses ($235~\ns$ full width at half maximum, 30 dB more power than used in (a)) with variable frequencies $\fa,\fb$ were then applied, followed by a final readout pulse. 
    The observed peak in the final $\oa$ population lies along $\fa = \fb + 609~\MHz$ (black dashed line).
    (c)
    Full $\Gamma$ histograms of the final readout pulse for the two subsets of measurements enclosed by the gray and black solid lines in (b).
    Data accrued in the region enclosed by the gray line shows little population transfer from the post-selected $\ob$ (left panel), while data in the region enclosed by the black shows significant population transfer to $\oa$ (right panel).
    \label{fig2}
	}
\end{figure}

The first step in driving the Raman process [Fig. 1(b)] was locating the two transitions that defined the $\Lambda$ system: $\oa \leftrightarrow \oC$ and $\ob \leftrightarrow \oC$. 
After breaking spin degeneracy with $\Phi = -0.10\Phi_0$ and tuning the transitions to a local maximum in $\Vg$ to mitigate charge noise (see Supplementary Information), we measured the spectrum shown in Fig.\ref{fig2}(a) using two-tone spectroscopy, without spin initialization. 
The dip in $P_\uparrow$ at $13.000~\GHz$ corresponds to the drive coming into resonance with the $\ob \leftrightarrow \oC$ transition, resulting in population transfer out of $\ob$ and into $\oC$. 
Similarly, the dip in $P_\downarrow$ at $13.684~\GHz$ corresponds to the $\oa \leftrightarrow \oC$ transition (see Supplementary Information for gate voltage and $\Phi$ dependence). 
Taking the difference yields the spin splitting $\es/h = 684~\MHz$. 

Having characterized the $\Lambda$ system, we then investigated two-photon Raman transitions of the trapped quasiparticle. 
After initializing the quasiparticle in $\ob$, we applied two simultaneous Gaussian pulses with variable respective carrier frequencies $\fb$ and $\fa$ and then measured the final qubit spin state [Fig.~\ref{fig2}(b)].
Throughout the main text, we present data with $\fb$ and $\fa$ blue-detuned from $\ob \leftrightarrow \oC$ and $\oa \leftrightarrow \oC$ respectively (see Supplementary Information for data over a wider frequency range). 
Along a line given by $\fa = \fb + 609~\MHz$, we observe increased $\oa$ population that we attribute to the onset of a Raman process. 
As expected for Raman transitions, the slope of this line is equal to one, since a shift of one drive frequency must be compensated by an equal shift of the other. 
The discrepancy between the spin splitting $\es/h = 684~\MHz$ and the $609~\MHz$ offset was due to an uncontrolled shift of the Andreev spectrum that occurred in between the measurements shown in Fig.~\ref{fig2}(a) and Fig.~\ref{fig2}(b) (see Supplementary Information).   

To further illustrate the dynamics of the quasiparticle under the Raman transitions, we histogram $\Gamma$ for data points off/on resonance with the Raman process [Fig.~\ref{fig2}(c)]. 
Off resonance, the quasiparticle was found predominantly in $\ob$, as expected from post-selection on the initial readout pulse.
On resonance, there was significant population transfer to $\oa$ as desired, as well as a small population transfer to $\g$. 
This is due to drive-induced quasiparticle de-trapping, which we comment on further below. 

\begin{figure}[h]
	\includegraphics[width=\columnwidth]{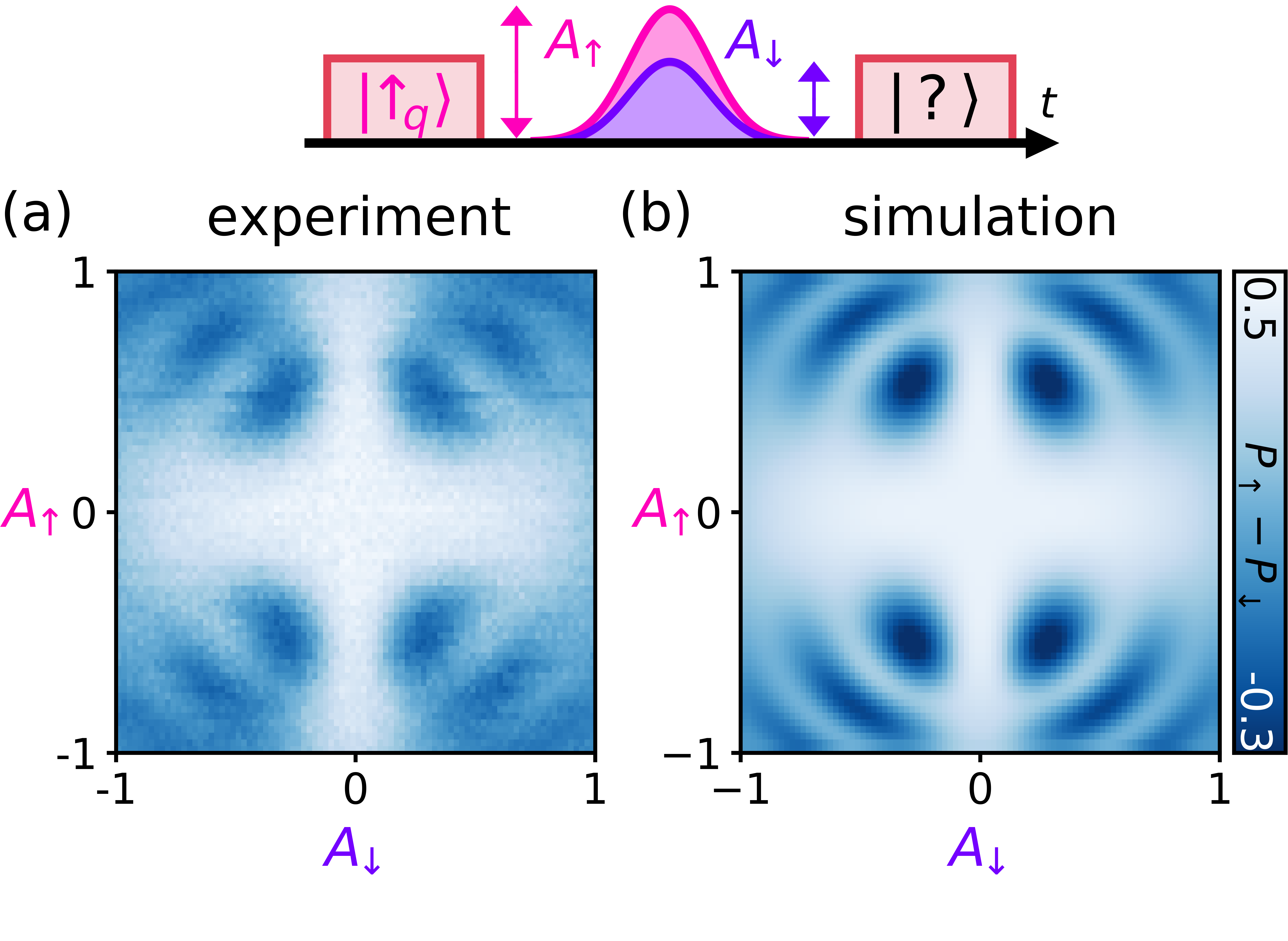} 
	\caption{
    Coherent $\Lambda$-Rabi oscillations of the quasiparticle spin at $\fa=13.280~\GHz$ and $\fb=13.964~\GHz$. 
    (a)
    Independently varying the amplitudes $A_\uparrow$ and $A_\downarrow$ of the simultaneous Gaussian drive pulses ($94~\ns$ full width at half maximum) resulted in coherent oscillations between $\ob$ and $\oa$ characteristic of a Raman process. 
    The oscillations are only present away from $A_\uparrow,A_\downarrow = 0$, and are symmetric under sign flips of $A_\downarrow,A_\uparrow$. 
    (b) 
    Simulated dynamics of the quasiparticle under the action of the drive pulse. 
    The reduced contrast observed in (a) is taken into account using the measured readout fidelities. 
    \label{fig3}
	}
\end{figure}

Finally, having induced spin population transfer using Raman transitions, we demonstrate the first coherent manipulation of the spin of an individual quasiparticle.
We first chose our Raman drive frequencies using the same measurement as shown in Fig. 2, but with shorter pulses ($94~\ns$ full width at half maximum): we detuned $\fa$ by $280~\MHz$ from $\oa \leftrightarrow \oC$ and varied $\fb$ until we observed maximum spin population transfer. 
We then varied the amplitudes $A_\uparrow,A_\downarrow$ of the two Gaussian pulses before determining the final quasiparticle state [Fig.~\ref{fig3}(a)]. 
The observed oscillations in the population difference between the two spin states are characteristic of a coherent Raman process. 
Qualitatively, when either $A_\downarrow=0$ or $A_\uparrow=0$ there is no population transfer because both drives are required to induce the Raman process. 
As the amplitudes of both drives are increased (roughly along the diagonals $|A_\downarrow| \simeq |A_\uparrow|$), the spin population difference undergoes coherent oscillations. 
As expected, the data are symmetric under $A_\downarrow \rightarrow - A_\downarrow$ and $A_\uparrow \rightarrow - A_\uparrow$. 

Quantitatively, the data are well-represented by a simulation of the coherent quasiparticle dynamics under the action of the drive pulses [Fig.~\ref{fig3}(b)].
Details of this numerical simulation can be found in the Supplementary Information. 
Using a Lindblad master equation~\cite{johansson2013qutip} we calculated the dynamics of the quasiparticle between the two Andreev doublets, with the inter-doublet transition frequencies and dephasing rates, spin dephasing rate, state-dependent readout fidelities, and pulse frequencies and envelopes fixed to values determined by independent measurements and instrument settings.
We then fit the simulation to the measured data by varying the four inter-doublet transition matrix elements [Fig. 1(d)], as well as a slight detuning from the Raman resonance condition, which we found to be $5.5\pm0.1~\MHz$. 
We also included a phenomenological drive-induced quasiparticle de-trapping rate ($10.8\pm0.9~\MHz$ at $|A_\downarrow|= |A_\uparrow|=1$) to capture the measured increase of $\oa, \ob \rightarrow \ket{g}$ for larger drive powers. 
We were thus able to capture the measured Raman spin dynamics of a quasiparticle trapped in the Andreev levels of a Josephson nanowire. 

\begin{figure}[h]
	\includegraphics[width=\columnwidth]{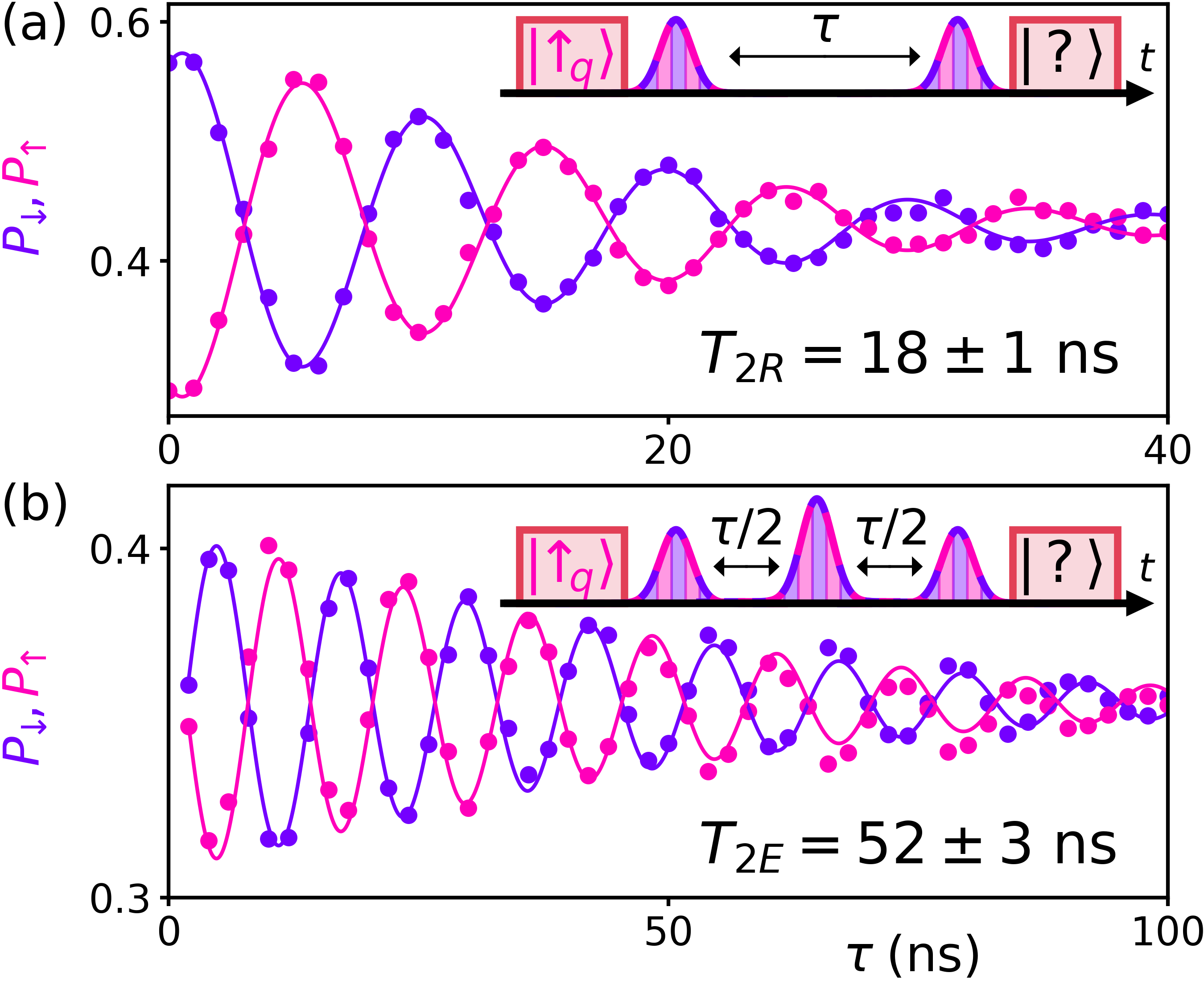} 
	\caption{
    Coherence decay of the quasiparticle spin ($\Vg = -59.1~\mV,~\Vnw = -33.3~\mV,\Phi = -0.115\Phi_0$).
    Ramsey (a) and Hahn-echo (b) experiments reveal $\TR = 18 \pm 1~\ns$ and  $\TE = 52 \pm 3~\ns$, respectively.
    Solid lines indicate fits to the data (see main text). 
    Oscillations were introduced in both cases by adding a phase proportional to $\tau$ to the final Raman pulse. 
    \label{fig4}
	}
\end{figure}

With the ability to perform coherent spin manipulation in hand, we then characterized the coherence lifetime of an Andreev spin.
A Ramsey measurement [Fig.~\ref{fig4}(a)] revealed spin coherence decay with a timescale $\TR = 18 \pm 1 ~\ns$, while a Hahn-echo pulse sequence [Fig.~\ref{fig4}(b)] resulted in a slightly longer timescale $\TE = 52 \pm 3 ~\ns$. 
Both measurements were well-described by a decay envelope $\exp{[-(\tau/T_2)^{1+\alpha}]}$ with $\alpha=~0.3\pm0.1$, indicative of excess low-frequency components compared to a white noise spectrum where $\alpha$ would be zero.   
The observed oscillations in $P_\uparrow, P_\downarrow$ are centered about a lower value in Fig.~\ref{fig4}(b) as compared to~\ref{fig4}(a), which we attribute to additional quasiparticle de-trapping $\oa, \ob \rightarrow \ket{g}$ caused by the echo pulse. 
Both the observed Ramsey and Hahn-echo coherence times are comparable to that of the spin-orbit qubit~\cite{nadj2010spin, petersson2012circuit, van2013fast}, the closest cousin of the Andreev spin qubit in that it consists of the spin-orbit hybridized pseudospin of a single electron. 
However, because here the quasiparticle was trapped in Andreev levels, we possessed a different experimental lens with which to investigate the effects of the environment on the spin coherence. 

As the Andreev levels of a Josephson nanowire are tunable via both electrostatic voltages and flux, we first suspected charge or flux noise as the source limiting the Andreev spin qubit coherence. 
However, we found that neither $T_{2R}$ nor $T_{2E}$ varied with $\Vg$ around the sweet-spot bias point (see Supplementary Information).
This indicated that the spin coherence was not limited by charge noise, consistent with the observed weak dependence of $\es$ on $\Vg$.
By comparing to the charge-noise-limited inter-doublet transitions, we extracted a lower-bound of $4.2~\us$ on the charge-noise-induced dephasing time of the quasiparticle spin (see Supplementary Information).
Moreover, the spin coherence time was not limited by flux noise, as we found no measurable dependence on $\Phi$. 

To better understand what was limiting the coherence of the Andreev levels, we additionally measured the coherence lifetimes of both inter-doublet transitions and so-called ``pair transitions'' at several gate bias points [Tab. 1].
The latter correspond to the excitation of two quasiparticles out of the condensate into both levels of a doublet~\cite{bretheau2013exciting, Janvier15, hays2018direct, tosi2019spin}. 
We found that the pair transition coherence times were systematically an order of magnitude longer than inter-doublet transition coherence times.
To first order, perturbations that couple to spin (such as a Zeeman field) result in equal and opposite energy shifts of the two doublet levels~\cite{tosi2019spin}.
As such, these perturbations do not change the frequency of the doublet pair transition, and therefore do not cause dephasing.
However, such spin-specific perturbations do induce dephasing of both inter-doublet transitions and the Andreev spin qubit.
It thus appears that the coherence lifetime of the Andreev spin qubit is limited by a spin-specific noise source such as hyperfine interactions with the spinful nuclei of indium and arsenic (though nuclear baths are typically lower frequency than the measured ratio $T_{2E}/T_{2R} = 2.9$ and decay envelope would indicate ~\cite{malinowski2017notch}), phonon-induced fluctuations of the nanowire spin-orbit coupling, or noisy paramagnetic impurities on the surface of the nanowire ~\cite{hanson2007spins}.

\begin{table}[h]
\centering
\begin{tabular}{ |c|c|c|c|c| } 
 \hline
 \textbf{Transition} & $\boldsymbol{\Vg~(\mV)}$ & $\boldsymbol{\Vnw~(\mV)}$ & $\boldsymbol{\TR~(\ns)}$ & $\boldsymbol{\TE~(\ns)}$ \\ 
 \hline
 $\ob \leftrightarrow \oC$ & -166.3 & -127.6 & n.m. & $39\pm8$ \\ 
 \hline
 pair & -164.9 & -127.6 & $38\pm9$ & $257\pm9$ \\ 
 \hline
 $\oa \leftrightarrow \oC$ & 144.5 & 24.9 & n.m. & $9\pm1$ \\ 
 \hline
 pair & 116.2 & 23.8 & $11\pm1$ & $420\pm20$ \\ 
 \hline
 $\ob \leftrightarrow \oC$ & 32.7 & -4.7 & n.m. & $11\pm3$ \\ 
 \hline
 pair & 30.0 & -4.7 & $14\pm2$ & $490\pm10$ \\ 
 \hline
\end{tabular}
 \caption{
Coherence lifetimes of Andreev transitions at various bias points. Ramsey coherence times for the inter-doublet transitions were not measured (n.m.) as they were too short.}.
\end{table}

In this work, we have demonstrated the first coherent manipulation of an Andreev spin qubit by driving Raman transitions of a single quasiparticle spin. 
In future experiments, the fidelity of this process may be improved by the implementation of a resonant STIRAP protocol~\cite{cerrillo2020}, which would reduce the necessary pulse amplitude and length, thereby mitigating the effects of dephasing and quasiparticle de-trapping. 
The parity dynamics may also be suppressed by improved filtering of high-frequency radiation~\cite{serniak2019direct}. 
In addition, engineering of the nanowire mirror symmetry could result in larger spin-flip drive matrix elements without sacrificing the spin lifetime.
Critical to the demonstration of spin manipulation was our ability to perform single-shot, cQED readout, which inherently relies on substantial coupling between the Josephson nanowire and a microwave resonator. 
In future experiments, this coupling could be used to achieve long-range interaction between separate qubits, a worthy goal in spin qubit research~\cite{petersson2012circuit, mi2018coherent, samkharadze2018strong, landig2018coherent, borjans2020resonant}. 
The nature of the spinful bath limiting the quasiparticle spin coherence may be elucidated by the application of magnetic field~\cite{hanson2007spins}, which could also push the Josephson nanowire through a topological phase transition~\cite{fu2008superconducting,Lutchyn2010, Oreg2010, Mourik2012, deng2016majorana}. 
In general, a detailed understanding of the effect of magnetic field on quasiparticle dynamics will be critical for further progress.

We thank Gijs de Lange for assistance with device design, and thank Nick Frattini and Vladimir Sivak for providing us with a SNAIL parametric amplifier.
We are grateful to Marcelo Goffman, Cyril Metzger, Hugues Pothier, Leandro Tosi, and Cristián Urbina for sharing their experimental results and hypotheses. 
We acknowledge useful discussions with Nick Frattini, Luigi Frunzio, Leonid Glazman, Manuel Houzet, Pavel Kurilovich, Vlad Kurilovich,
 and Charles Marcus.

This research was supported by the US Office of Naval Research (N00014-16-1-2270) and by the US Army Research Office (W911NF-18-1-0020, W911NF-18-1-0212 and W911NF-16-1-0349). 
D.B. acknowledges support by Netherlands Organisation for Scientific Research (NWO) and Microsoft Corporation Station Q.
J.C.  acknowledges the  support  from MICINN (Spain) (“Beatriz  Galindo”  Fellowship BEAGAL18/00081).
J.N. acknowledges support from the Danish National Research Foundation. 
Some of the authors acknowledge the European Union’s Horizon 2020 research and innovation programme for financial support: A.G received funding from the European Research Council, grant no. 804988 (SiMS), and A.G., A.L.Y., J.C., and J.N. further acknowledge grant no. 828948 (AndQC) and QuantERA project no. 127900 (SuperTOP).
A.L.Y. acknowledges support by Spanish MICINN through grants FIS2017-84860-R and through the “María de Maeztu” Programme for Units of Excellence in R\&D (Grant No. MDM-2014-0377).

\noindent \textbf{Contributions}

\noindent M.H., V.F., K.S., D.B., T.C., A.G., and M.D. designed the experimental setup.
P.K. and J.N. developed the nanowire materials.
D.B. and A.G. fabricated the device.
M.H. and V.F. performed the measurements. 
V.F., M.H., J.C. and A.L.Y. developed the symmetry analysis and microscopic modeling.
M.H., J.C., V.F., and A.L.Y. developed and performed the Raman simulations.
M.H., V.F., K.S., S.D., and M.D. analyzed the data.
M.H., V.F., and M.D. wrote the manuscript with feedback from all authors.


\clearpage
\onecolumngrid
\begin{center}
{\Large \textbf{Supplementary Information}}
\end{center}


\setcounter{equation}{0}
\setcounter{figure}{0}
\setcounter{table}{0}
\setcounter{page}{1}
\makeatletter
\renewcommand{\theequation}{\arabic{equation}}
\renewcommand{\thefigure}{S\arabic{figure}}

\section{I. Symmetry Analysis Of Pseudospin-Flip Matrix Elements}

\vspace{-2mm}
In this section we motivate a simple physical picture to explain the qualitative features of the driven spin-flipping transitions. 
 First, note that while an electron spin does in principle couple to the magnetic field of our microwave drive, this coupling is extremely weak. 
The electric field of the drive, on the other hand, only couples to motional degrees of freedom, so it cannot induce spin-flip transitions if spin is a good quantum number.
Spin-orbit coupling is the conventional invocation to solve this problem since it hybridizes spin and translational wavefunction components into what is often referred to as pseudospin~\cite{nadj2010spin}.
We will use symmetry considerations to show that this conversion to pseudospin is necessary but insufficient to allow electric fields to flip the quasiparticle pseudospin in our system, and that an additional broken spatial symmetry is required.
Following this we further validate these ideas by inspection of a tight-binding model that specifically incorporates the physics and energy scales of Andreev levels. 
\vspace{-2mm}
\subsection{A. General Considerations}
\vspace{-2mm}

We are interested in how transitions between the Andreev levels of the Josephson nanowire are induced by our microwave drive voltage $V_d \cos \omega_d t$.
Here $\omega_d$ is the drive frequency and $V_d$ the spatial profile. 
Initially, we will model the spatial profile as a purely longitudinal differential voltage along the Josephson nanowire $V_{d}(x) \approx -V_{d}(-x)$, which is a reasonable starting point given our highly symmetric device design [Fig.~\ref{fig_wiring_diagram}].
The transition rates will depend both on the matrix element of $V_d$ between the initial/final states, as well as on the mismatch between $\hbar \omega_d$ and the energy difference between the initial/final states.  
We first focus on matrix element considerations, initially assuming that $\Phi = 0$ such that the system is time-reversal symmetric, before generalizing to any value of $\Phi$. 

Let's begin by imagining the Josephson nanowire as a quasi-1d system described by a Hamiltonian $H_0$. 
This Hamiltonian includes both the superconducting leads and the semiconductor nanowire, though for the moment we neglect spin-orbit coupling.
Additionally, let us suppose that the system is rotationally invariant about the $x$-axis.
As with any spin-1/2 system with time-reversal symmetry, the energy levels are paired into spin-degenerate doublets (Kramers theorem). 
In order to achieve the Raman process investigated in this experiment, it was necessary to simultaneously drive  spin-conserving and spin-flipping inter-doublet transitions.  
Under our current model of the system, while $V_d$ can induce spin-conserving inter-doublet transitions by coupling to the spatial character of the wavefunctions, it cannot induce spin-flipping transitions because both $H_0$ and $V_d$ are block-diagonal in spin (they are spin-rotation-symmetric). 

Spin-orbit coupling can help solve this problem by mixing spin and motional degrees of freedom.
We can see this in the form of a Rashba interaction generated by a static electric field in the $z$ direction $H_R = i E_z \gamma (\sigma_x \partial_y - \sigma_y \partial_x)$, where $\sigma_i$ are Pauli matrices of the spin and $\gamma$ is a material parameter.
Thus, in the presence of spin-orbit interaction, the quasiparticle ``spin'' is actually a pseudospin.
We stress that a hybridization like this must be present in our system, as it is critical to break the Andreev-doublet degeneracy~\cite{governale2002spin, reynoso2012spin, tosi2019spin}. 
Upon including a Rashba interaction, the Hamiltonian $H = H_0 + H_R$ is no longer block-diagonal in spin, and one might imagine that $V_d$ could flip pseudospin. 

However, a selection rule prevents this.
While the static electric field in the $z$ direction associated with the Rashba effect breaks rotational symmetry, a transverse mirror symmetry in the $y$ direction remains [see main text Fig. 1(c)]. 
This mirror symmetry is described by the operator $M_y=-i\sigma_y \delta_{y,-y}$, where $\delta_{y,-y}$ sends $y$ to $-y$. 
Because $[H, M_y]=0$, the energy eigenstates are also mirror eigenstates. 
Moreover, as spin is flipped under time-reversal $T \sigma_y T^{-1} = -\sigma_y$, the mirror eigenvalue is also flipped. 
Therefore, the pair of pseudospins comprising each doublet have mirror eigenvalues $+i$ and $-i$ (this remains true in the presence of a time-reversal-symmetry-breaking phase-bias, as explained below).
As our longitudinal drive $V_d(x) \approx -V_d(-x)$ also obeys the mirror symmetry $[V_d, M_y]=0$, it cannot induce transitions between states of different mirror eigenvalue (pseudospin flips).

To induce pseudospin-flip transitions, the mirror symmetry must be broken. 
In the real device, this symmetry is broken by the epitaxial aluminum shell which covers two of six nanowire facets, the presence of the side gates and their applied voltages [main text Fig. 1(c)], as well as any non-idealities of the device.  
As such, the symmetry may be broken in any/all of the Hamiltonian terms:

\begin{itemize}
    \item $[H_0, M_y]\neq0$: asymmetric superconducting leads, gate-electrode perturbations to the transverse confinement
    \item $[H_R, M_y]\neq0$: modified Rashba interaction due to non-symmetric electric field profiles
    \item $[V_d, M_y]\neq0$: drive applied via asymmetric leads and perturbed by presence of the metallic gate
\end{itemize}

\noindent Under this broken mirror symmetry, both the pseudospin-conserving transition $\ob \leftrightarrow \oC$ and the pseudospin-flipping transition $\oa \leftrightarrow \oC$ are allowed and therefore the Raman process can be driven. 

Thus far in this section, we have assumed $\Phi=0$ such that $T H T^{-1} = H$. 
Because the drive also obeys time-reversal symmetry $T V_d T^{-1} = V_d$, the direct pseudo-spin transition $\oa \leftrightarrow \ob$ is thus forbidden for $\Phi=0$. 
While this feature is ideal for the use of the Andreev doublets as a $\Lambda$ system, in this experiment it was necessary to break the doublet degeneracy such that $\oa \leftrightarrow \oC$ and $\ob \leftrightarrow \oC$ were frequency-resolved. 
We therefore performed most measurements at a nonzero flux bias ($\Phi \approx -0.1\Phi_0$) such that we could drive the inter-doublet transitions as well as detect the pseudospin state of the lower doublet. 
As we investigate numerically in the next section, we expect the direct spin-flip transition to remain suppressed even for nonzero $\Phi$.
Note that as long as the additional time-reversal-breaking terms in $H$ respect the mirror symmetry, as is the case for a dc phase applied to mirror-symmetric leads, all the inter-doublet transition considerations presented above for $\Phi = 0$ automatically generalize to the case of nonzero $\Phi$.
In particular, one state of each doublet has mirror eigenvalue $+i$ and the other $-i$, independent of $\Phi$. 

\subsection{B. Numerical Tight-Binding Model} \label{supp_tight_binding}

We now explore these symmetry considerations numerically using a tight-binding model. We begin by investigating spin-orbit hybridization and broken mirror symmetry in the band structure of an un-proximitized nanowire, before moving on to the Andreev levels themselves. 
\vspace{-4mm}
\subsubsection{1. Unconfined Normal Channels}
\begin{figure}[H]
    \centering
	\includegraphics[width=4in]{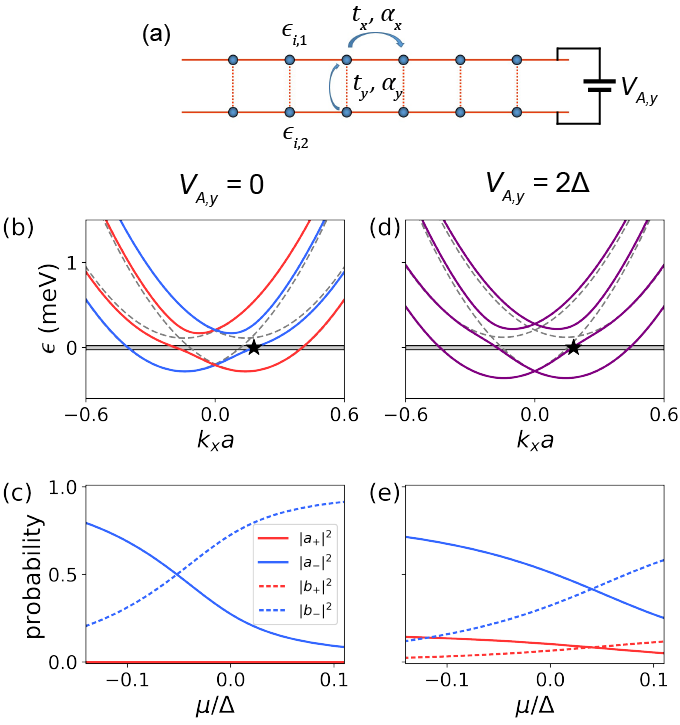} 
	\caption{\label{fig_tight_binding}
	(a) Visual representation of the tight-binding model of the un-proximitized infinite nanowire.
	(b) Band structure of the tight-binding model for $V_{A,y} = 0$. Bands split by the longitudinal Rashba term (gray dashed lines) undergo avoided crossings due to the transverse Rasbha term to give the final bands of the model (colored lines). Bands colored in blue are anti-symmetric eigenstates of $M_y$, and bands colored in red are symmetric eigenstates. 
	The black star marks the slow, positive momentum Fermi point, the transverse wavefunction components of which are plotted in (c) for the values of $\mu$ indicated by the gray strip.
	Only the anti-symmetric amplitudes $a_-$ and $b_-$ are nonzero.
	As the chemical potential is tuned through the avoided crossing between the bands, weight shifts from $\ket{\Leftarrow,S}$ to $\ket{\Rightarrow,A}$ (see Eqn. S3).
	(d) For $V_{A,y} = 2 \Delta$, the bands are qualitatively similar to (b), but are no longer mirror eigenstates. 
	As shown in (e), the wavefunction of the slow, positive-momentum Fermi point has both symmetric and anti-symmetric components.
	}
\end{figure}

\vspace{-2mm}
Our model of the un-proximitized nanowire is graphically represented in Fig.~\ref{fig_tight_binding}(a). 
It consists of two coupled infinite parallel chains aligned along the $x$-direction, and with Rashba spin-orbit coupling: 
\vspace{-2mm}
\begin{eqnarray}\label{two_channel_normal}
H &= &\sum_{i,\tau,\sigma} (\epsilon_{i,\tau} - \mu)c^{\dagger}_{i,\tau,\sigma}c_{i,\tau,\sigma}
+ t_x c^{\dagger}_{i,\tau,\sigma} c_{i+1,\tau,\sigma} + \sigma \alpha_x c^{\dagger}_{i,\tau,\sigma} c_{i+1,\tau,\bar{\sigma}} + \mbox{H.c.} \nonumber\\
&&  + \sum_{i,\sigma} t_y c^{\dagger}_{i,1,\sigma}c_{i,2,\sigma} + i\alpha_y c^{\dagger}_{i,1,\sigma}c_{i,2,\bar{\sigma}} + \mbox{H.c.}
\end{eqnarray} 
\vspace{-2mm}

\noindent Here the operators $c^{\dagger}_{i,\sigma,\tau}$ create an electron on the site $i$, within the channel $\tau=1,2$ and spin $\sigma$.
The scalar hopping strengths $\{t_x,t_y\}$ correspond to the  \{longitudinal, transverse\} directions, while the Rashba hopping strengths are given by $\{\alpha_x,\alpha_y\}$.
The strengths are scaled by the discretization of a continuous Hamiltonian with a longitudinal lattice parameter $a=20$ nm and width $W=100$ nm, giving $t_x = \hbar^2/(m^*a^2)$, $\alpha_x=\alpha/(2a)$, $t_y=\hbar^2/(m^*W^2)$ and $\alpha_y=\alpha/(2W)$, where $m^* = 0.028 m_e$ is the effective mass for InAs and $\alpha=E_z \gamma=25~\meV\cdot \nm$ is the Rashba parameter. 
In the mirror-symmetric case, the on-site energies $\epsilon_{i,\tau}$ are given by $2 t_x$. 
Momentarily, we will break this mirror symmetry by including an energy offset $V_{A,y}$ between the two chains such that $\epsilon_{1,\tau} \rightarrow \epsilon_{1,\tau} + V_{A,y}/2$ and $\epsilon_{2,\tau} \rightarrow \epsilon_{2,\tau} - V_{A,y}/2$. 
Below, we will express all energies as fractions of the superconducting gap of bulk aluminum $\Delta=0.185~\meV$.

\textcolor{white}{-}

\textcolor{white}{-}

\textcolor{white}{-}

\textcolor{white}{-}

\textcolor{white}{-}

\textcolor{white}{-}

\textcolor{white}{-}

\textcolor{white}{-}

\textcolor{white}{-}

\textcolor{white}{-}

\textcolor{white}{-}

\textcolor{white}{-}

\textcolor{white}{-}

\textcolor{white}{-}

\textcolor{white}{-}

\textcolor{white}{-}

\textcolor{white}{-}

\textcolor{white}{-}

\textcolor{white}{-}

\textcolor{white}{-}

\textcolor{white}{-}

\textcolor{white}{-}

\textcolor{white}{-}

\textcolor{white}{-}

First, we consider the mirror-symmetric case $V_{A,y}=0$.
The model has four bands [Fig.~\ref{fig_tight_binding}(b)]: the spatial character of the transverse wavefunction can either be symmetric $\ket{S}$ (lower energy) or anti-symmetric $\ket{A}$ (higher energy), while the spin can be either $\ket{\Rightarrow}$ or $\ket{\Leftarrow}$ in the $y$-direction. 
The longitudinal Rashba term $i\alpha_x c^{\dagger}_{i,\tau,\sigma} c_{i+1,\tau,\bar{\sigma}} + \mbox{H.c.}$ generates momentum split-bands [gray dashed lines in Fig. S1(b)], but spin in the $y$-direction remains a good quantum number. 
However, the transverse part of the Rashba term $i \alpha \sigma_x \partial_y \rightarrow i\alpha_y c^{\dagger}_{i,1,\sigma}c_{i,2,\bar{\sigma}} + \mbox{H.c.}$ generates an avoided crossing between the lower-energy, transverse-symmetric band and the higher-energy, transverse-anti-symmetric band. 
This inter-band mixing results in hybridization between spin and transverse motional degrees of freedom, so that the transverse character of the resultant low-energy band is given by 

\begin{align}
     \ket{+}  &= a_{+} \ket{\Rightarrow,S} + b_{+} \ket{\Leftarrow,A} \\
     \ket{-}  &= a_{-} \ket{\Leftarrow,S} + b_{-} \ket{\Rightarrow,A} 
\end{align}

\noindent The numerically-calculated value for the $\ket{-}$ coefficients is shown in Fig.~\ref{fig_tight_binding}(c). 
As $\mu$ is swept through the avoided crossing, weight shifts from $\ket{\Leftarrow,S}$ to $\ket{\Rightarrow,A}$, with maximal hybridization occurring around $\mu \approx -0.05 \Delta$. 
Critically, although there is hybridization, the new bands are still eigenstates of the mirror operator $M_y \ket{\pm} = \pm i\ket{\pm}$. 
As such, a mirror-symmetric drive cannot induce transitions between them. 

To model a broken mirror symmetry, we set a non-zero inter-chain potential difference $V_{A,y} = 2\Delta$. 
The bands for this case are depicted in Fig.~\ref{fig_tight_binding}(d).
While the change in the energy dispersion relation is subtle, we can see the effect of the broken mirror symmetry in the wavefunction character [Fig.~\ref{fig_tight_binding}(e)]. 
The band near the avoided crossing now has character of all four basis states $\ket{\Rightarrow,S}, \ket{\Leftarrow,S}, \ket{\Rightarrow,A}, \ket{\Leftarrow,A}$. 
There are thus no selection rules forbidding transitions induced by a mirror-symmetric drive. 

\subsubsection{2. Andreev Levels}

\begin{figure}[H]
    \centering
	\includegraphics[width=5in]{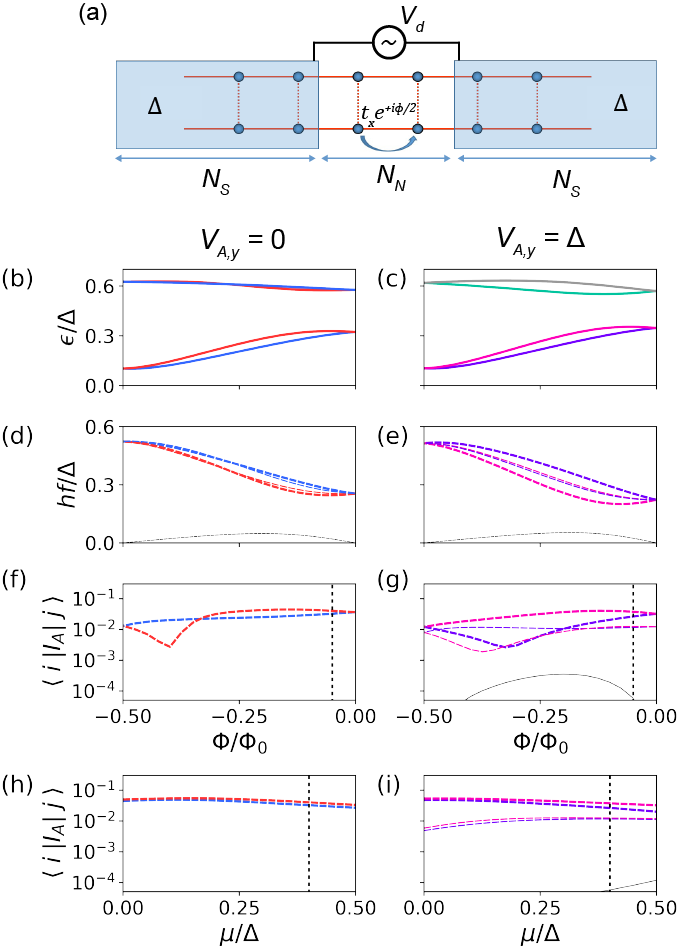} 
	\caption{\label{fig_andreev}
	(a) Visual representation of the tight-binding model, now including superconducting leads.
	(b-h) 
	For $V_{A,y} = 0,\Delta$, we compute the Andreev level energies (b,c), all transition frequencies (d,e), all $V_d$ matrix elements versus $\Phi$ (f,g), and $V_d$ matrix elements (drive amplitude $V_0 = 2\Delta$) versus $\mu$ involving the lowest energy level (h,i).
	For $V_{A,y} = 0$, states are colored by their mirror eigenvalue as before: red for $+1$ and blue for $-1$.
	For $V_{A,y} = \Delta$, states are colored the same as in main text Fig. 1(a).
	Inter-doublet transition frequencies and matrix elements take the same color as the participating lower doublet state, while those for the intra-doublet transition are grey.
	Thick dashed lines correspond to pseudospin-conserving transitions, while thin dashed lines correspond to pseudospin-flipping transitions. 
	Vertical dashed black lines in (f)/(g) correspond to $\Phi$ for (h)/(i), while vertical dashed black lines in (h)/(i) correspond to $\mu$ for (f)/(g).
	}
\end{figure}

Now we confine the normal region between two superconducting leads with a pair potential $\Delta$ and a different chemical potential $\mu_S$ [Fig. \ref{fig_andreev}(a)]. 
The total number of sites in each of the two chains is $N$. 
The superconducting phase difference is applied within the gauge of the $t_x$ hopping elements in the middle of the chain, i.e. between sites $N/2$ and $N/2+1$. 
Symbolically the Hamiltonian is 
\begin{eqnarray}\label{two-channel-model}
H &= &\sum_{i,\tau,\sigma} (\epsilon_{i,\tau} - \mu_i)c^{\dagger}_{i,\tau,\sigma}c_{i,\tau,\sigma}
+ t_x c^{\dagger}_{i,\tau,\sigma} c_{i+1,\tau,\sigma} - \alpha_x c^{\dagger}_{i,\tau,\sigma} c_{i+1,\tau,\bar{\sigma}} + \mbox{H.c.} \nonumber\\
&& + \sum_{i,\tau} \Delta_{i} c_{i,\tau,\downarrow} c_{i,\tau,\uparrow} + \mbox{H.c.} + \sum_{i,\sigma} t_y c^{\dagger}_{i,1,\sigma}c_{i,2,\sigma} + i\alpha_y c^{\dagger}_{i,1,\sigma}c_{i,2,\bar{\sigma}} + \mbox{H.c.}
\end{eqnarray} 

\noindent where $\mu_i=\mu_S$ for the sites in the superconducting leads and $\mu_i=\mu$ for the normal region as in the previous section.
We simulate the Hamiltonian Eqn. S\eqref{two-channel-model} in Nambu space, fixing $\mu_S=1.5\Delta$, $N = 34$, and the number of sites in each lead $N_\mathrm{leads}=6$.
We have found that while $N_\mathrm{leads}=6$ is large enough for the qualitative investigation we present here, more lead sites are necessary if quantitative accuracy is desired. 

We examine both the mirror-symmetric case $V_{A,y} = 0$ and the non-symmetric case $V_{A,y} = \Delta$. 
We note that we apply $V_{A,y}$ only to the sites in the normal region $N_\mathrm{leads} < i < N - N_\mathrm{leads}$. 
For both the mirror-symmetric and non-symmetric cases, we find Andreev levels with qualitatively similar energies  [Fig. \ref{fig_andreev}(b/c)] and therefore similar inter-doublet transition frequencies [Fig. \ref{fig_andreev}(d/e)].
For $V_{A,y}=0$, the Andreev levels are eigenstates of the mirror operator just as in the states of the infinite nanowire. 

Next, we check the drive matrix elements. 
We model the drive voltage profile as a linear potential drop between the leads $V_d(x) = \frac{2}{L} V_0 x$, choosing $V_0=2\Delta$.
For the mirror-symmetric case, only mirror-preserving transitions can be driven regardless of flux [Fig. \ref{fig_andreev}(f)] and chemical potential [Fig. \ref{fig_andreev}(h)], as expected. 
For $V_{A,y}=\Delta$, we indeed find that all transitions are allowed. 
However, while the inter-doublet transitions are finite for all values of $\Phi$, the direct spin-flip matrix element goes to zero at $\Phi \rightarrow 0, \Phi_0/2$ as required by time-reversal symmetry. 
We also find that the matrix element of pseudospin preserving and flipping transitions become of similar magnitude for higher chemical potentials [Fig. \ref{fig_andreev}(h)]. 
This is consistent with the stronger degree of spin-orbital mixing when the chemical potential is close to the next sub-band as shown in Fig.~\ref{fig_tight_binding}(e).

\textcolor{white}{-}

\textcolor{white}{-}

\textcolor{white}{-}

\textcolor{white}{-}

\textcolor{white}{-}

\textcolor{white}{-}

\textcolor{white}{-}

\textcolor{white}{-}

\textcolor{white}{-}

\textcolor{white}{-}

\textcolor{white}{-}

\textcolor{white}{-}

\textcolor{white}{-}

\textcolor{white}{-}

\textcolor{white}{-}

\textcolor{white}{-}

\textcolor{white}{-}

\textcolor{white}{-}

\textcolor{white}{-}

\textcolor{white}{-}

\textcolor{white}{-}

\textcolor{white}{-}

\textcolor{white}{-}

\textcolor{white}{-}

\textcolor{white}{-}

\textcolor{white}{-}

\textcolor{white}{-}

\section{II. Experiment schematic}

\begin{figure}[H]
	\centering
	\includegraphics[width=3in]{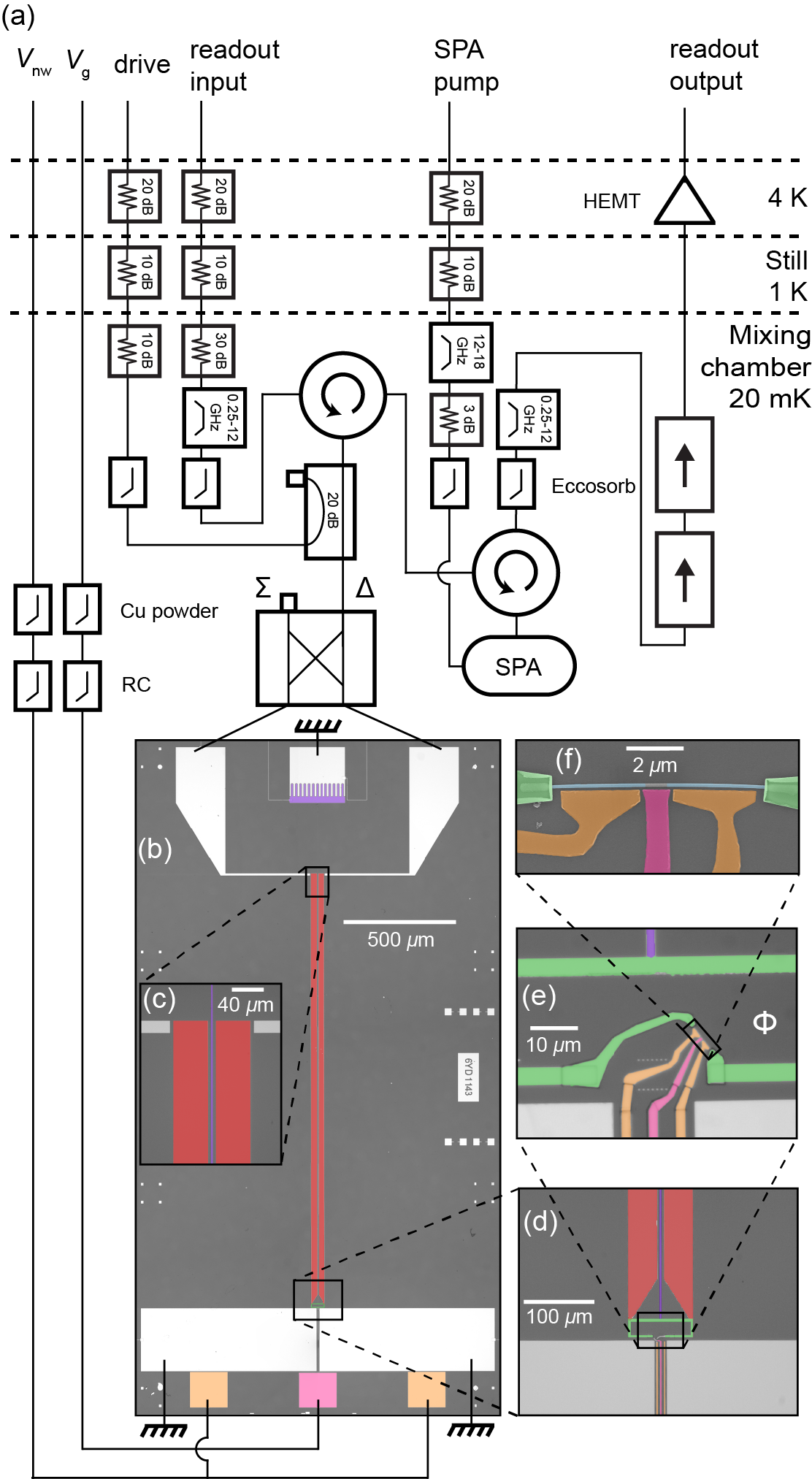} 
	\caption{\label{fig_wiring_diagram}
	Cryogenic wiring diagram and device micrographs (see Extended Data of Ref. ~\cite{hays2020continuous} for original publication).
    Optical micrograph (e) is of the device on which the presented measurements were performed. 
    Optical micrographs (b), (c), (d) and scanning electron micrograph (f) are of an extremely similar (unmeasured) device, the main difference being that the length of the weak length is $750~\nm$ instead of $500~\nm$.
    The microwave readout and drive tones pass through the depicted circuitry (a) before being routed through the $\Delta$ port of a $180^{\circ}$ hybrid resulting in differential microwave voltages at the device input. 
    After reaching two coupling capacitors (c), the readout tone was reflected off the differential $\sim\lambda/4$ mode of the coplanar strip resonator (red, frequency $\fr = 9.18843~\GHz$, coupling $\kappa_\mathrm{c} = 2\pi \times 1.23~\MHz$, internal loss $\kappa_\mathrm{i} = 2\pi \times  1.00~\MHz$) and then routed through the depicted amplification chain (a), which was comprised of a SNAIL parametric amplifier (SPA) \cite{frattini2018}, HEMT, and room-temperature amplifiers.  
    In this circuit, the drive tone creates an ac phase drop across the nanowire (f), which is embedded in the superconducting $\Phi$-bias loop (green) at the end of the resonator (d,e).
    One edge of the loop connects the two strips of the resonator and thereby forms the shared inductance with the nanowire.
    We controlled the electrostatic potential in the nanowire weak link (f) with a dc gate (pink, voltage $\Vg$).
    Gates on the nanowire leads (orange) were used to gain additional electrostatic control (voltage $V_\mathrm{nw}$).
    To reference the resonator/nanowire island to ground, an additional strip runs between the resonator strips, and connects to a large finger capacitor (purple). 
    This strip does not significantly perturb the resonator's microwave properties because it resides at the zero voltage point with respect to the resonator's differential mode.
    \label{device}
    }
\end{figure}

\section{III. Tuning up the device}

\begin{figure}[H]
    \centering
	\includegraphics[scale = 0.36]{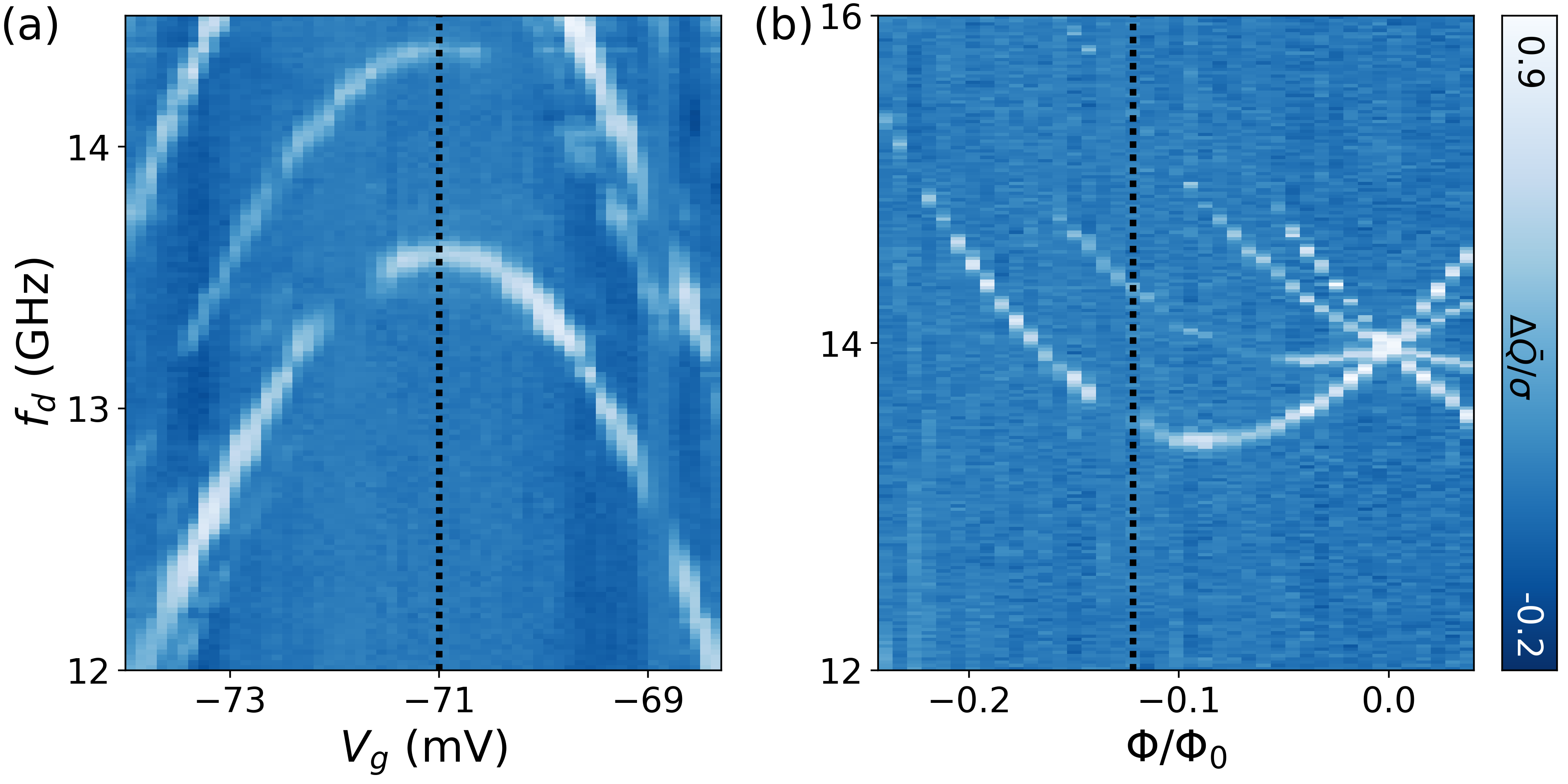} 
	\caption{Gate and flux dependence of the inter-doublet transitions. 
	(a) 
	A local maximum (``sweet spot'') is observed in both $\ob \leftrightarrow \oC$ and $\oa \leftrightarrow \oC$ at $\Vg = -71.0~\mV$ (black dotted line). 
	(b)
	Flux dependence of the four inter-doublet transitions at $\Vg = -71.0~\mV$, $\Vnw = 0.0~\mV$.
	Black dotted line indicates $\Phi$ bias for data shown in (a). 
	\label{finding_sweet_spot}
	}
\end{figure}

In this experiment, we possessed three \textit{in-situ} control knobs of the nanowire Andreev levels: a loop flux $\Phi$ [Fig. \ref{device}(d, e)], a main gate voltage $\Vg$ acting on the nanowire weak link [Fig. \ref{device}(f)], and an additional gate voltage $\Vnw$ applied to two more gates positioned on either side of the main gate [Fig. \ref{device}(f)]. 
Upon cooling down the device, we observed $\Phi$ and gate voltage dependence of the readout resonance around $\Vg, \Vnw = 0$, indicating that conduction channels in the nanowire link were transmitting. 
With $\Phi = -0.13\Phi_0$ and $\Vnw = 0$, we swept $\Vg$ while performing two-tone spectroscopy [Fig. \ref{finding_sweet_spot}(a)]. 
We observed several dispersing transitions, with a local maximum at $\Vg = -71.0~\mV$. 

Parking $\Vg$ at the local maximum to mitigate the effects of electrostatic noise on the Andreev level coherence (see below for further data and discussion), we then performed two-tone spectroscopy while sweeping $\Phi$ [Fig. \ref{finding_sweet_spot}(b)]. 
Four flux-dependent resonances were observed, which cross at $\Phi = 0$.  
This is characteristic of inter-doublet transitions of a quasiparticle between spin-orbit split Andreev levels~\cite{tosi2019spin}. 
In conjunction with the population transfer measurements shown in Fig. 2(a) of the main text,  this characteristic spectrum allowed us to identify the two lowest-frequency transitions as $\ob \leftrightarrow \oC$ and $\oa \leftrightarrow \oC$. 

\begin{figure}[H]
	\centering
	\includegraphics[scale = 0.36]{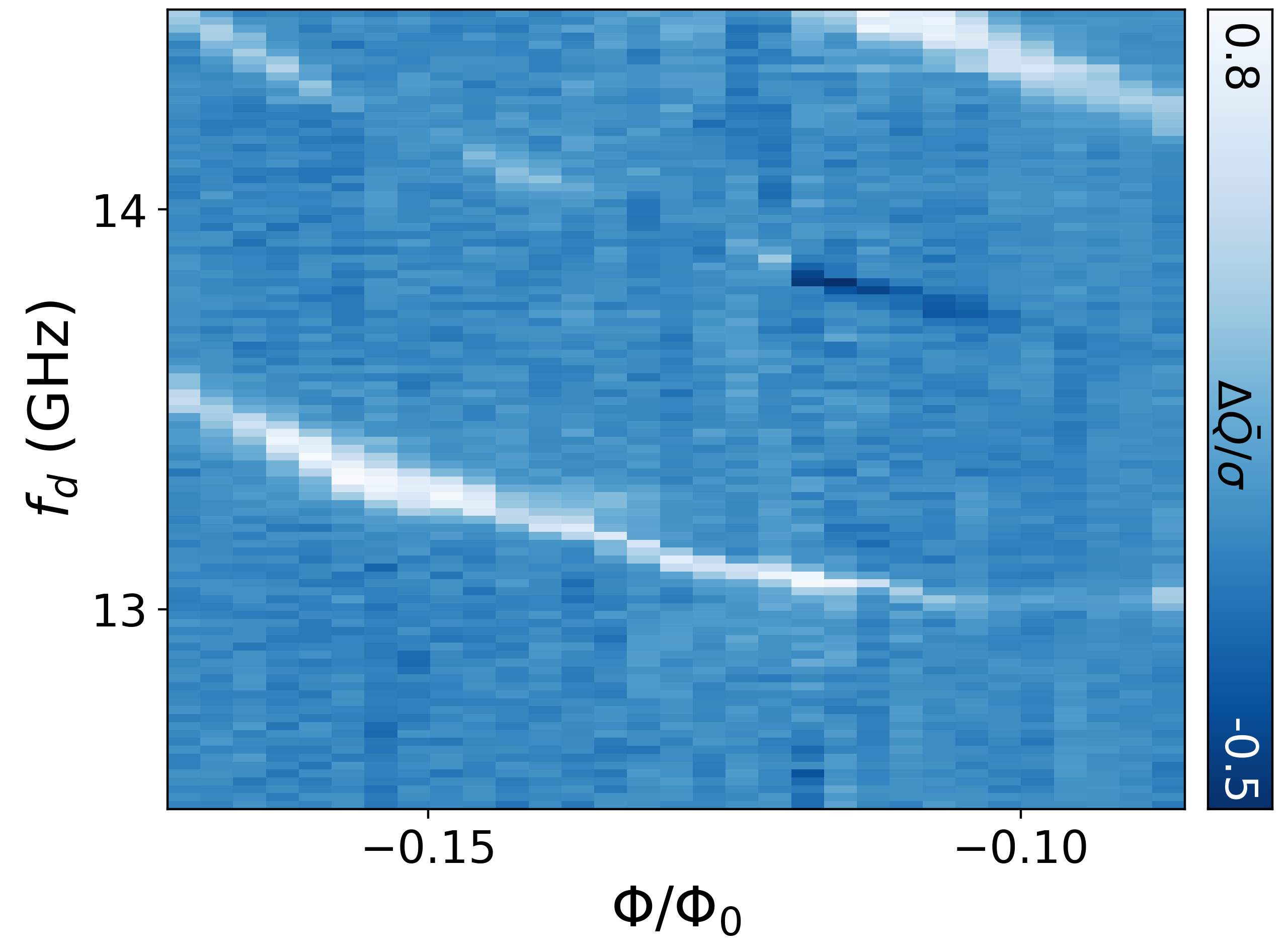} 
	\caption{Flux dependence of $\ob \leftrightarrow \oC$ and $\oa \leftrightarrow \oC$ at the bias point for all data presented in the main text, excluding Fig. 4 ($\Vg = -71.9~\mV$, $\Vnw = 4.0~\mV$).
	\label{sweet_spot_spec}
	}
\end{figure}

At certain $\Phi$ bias points, some of the transitions are not visible, or become significantly dimmer. 
For example, the $\ob \leftrightarrow \oC$ transition is barely visible at $\Phi =\simeq -0.13\Phi_0$.
This drop in signal occurs because the quasiparticle population of the relevant level of the lower doublet (either $\oa$ or $\ob$) decreases, often below 0.01. 
We attribute these population drops to evacuation of the quasiparticle into cold, dot-like levels in the nanowire that are brought into resonance with the Andreev doublets as $\Phi$, $\Vg$, and $\Vnw$ are varied. 
While these features are not completely understood, we found they could be easily avoided with an appropriate choice of bias conditions. 
The effects of these population drops can be observed in Figs. S\ref{finding_sweet_spot} and S\ref{sweet_spot_spec}. 

Having identified the transitions that defined the $\Lambda$ system, we searched for a local maximum of the transitions in both gate voltages in order to mitigate electrostatic noise. 
We found such a bias point at $\Vg = -71.9~\mV$ and $\Vnw = 4.0~\mV$ (see Fig. \ref{sweet_spot_spec} for $\Phi$-dependence at these gate voltages).

\section{IV. Searching for Raman transitions}

\begin{figure}[H]
	\centering
	\includegraphics[scale = 0.4]{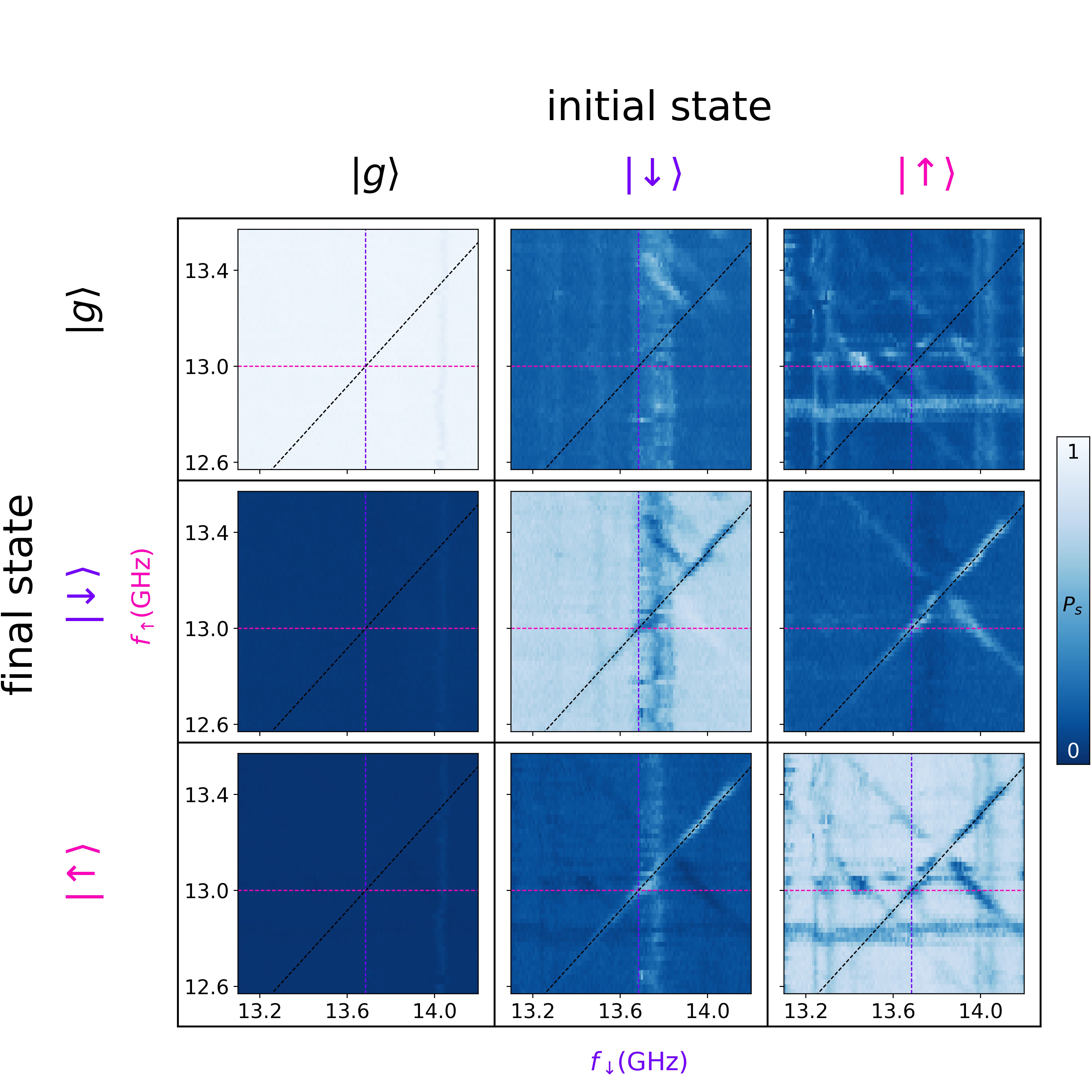} 
	\caption{
	All transition probabilities under the action of simultaneous drive pulses of variable carrier frequencies, as in Fig. 2(b) of the main text. 
	Measured transition frequencies of $\ob \leftrightarrow \oC$ and $\oa \leftrightarrow \oC$ are indicated by pink and purple dashed lines, respectively. 
	Black dashed lines have a slope of one, and run through the crossing point of the two transitions. 
	\label{all_raman}
	}
\end{figure}

With the gate voltages set to the optimum values discussed above and $\Phi = -0.10\Phi_0$, we applied simultaneous Gaussian drive pulses of variable carrier frequency, as explained in the main text and Fig. 2(b).
Here we present all measured transition probabilities, and over a wider frequency range than in the main text [Fig.~\ref{all_raman}]. 
The drive powers used in this measurement were 30 dB larger than in the two-tone spectroscopy measurement of Fig. 2(a).
The $\oa \leftrightarrow \oC$ transition is visible (though broadened) just above its low-power value, while the $\ob \leftrightarrow \oC$ transition is no longer visible. 
Other transitions that are constant in one drive frequency or the other are observed, perhaps due to evaporation of the quasiparticle into the same dot-like levels as discussed above.
Several multi-photon transitions are observed that lie along $\fb = -\fa + c$. 
The $-1$ slope indicates that the two drive frequencies are adding to reach a highly-excited state of the system, perhaps exciting the quasiparticle above the superconducting gap or into other dot-like states in the nanowire. 
Finally, the desired two-photon Raman process $\oa \leftrightarrow \ob$ occurs along the black dashed line, which has slope 1 and intersects with the crossing point of the two single-photon transitions of the $\Lambda$ system (unlike the data shown in Fig. 2(b) of the main text). 
This measurement was taken between the measurements displayed in Fig. 2(b) and 2(a) of the main text. 
We thus conclude that there was an uncontrolled shift of the Andreev levels between the measurement shown Fig.~\ref{all_raman} and Fig. 2(b), as referenced in the main text. 
Such jumps were not uncommon in this experiment, and occurred on a timescale of days to weeks. 

\section{V. Coherent spin dynamics}

\begin{figure}[H]
	\centering
	\includegraphics[scale = 0.12]{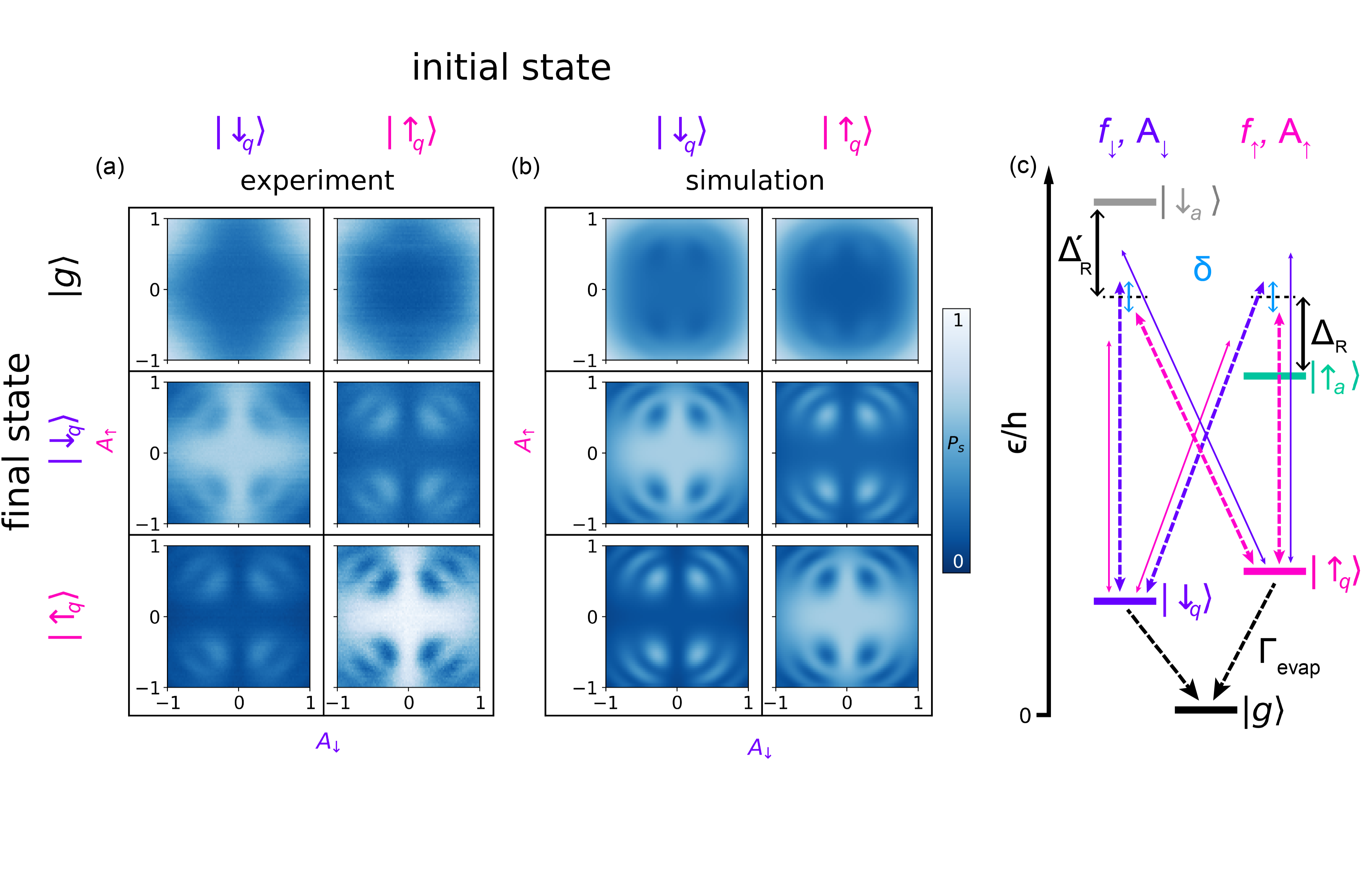} 
	\caption{Coherent $\Lambda$-Rabi oscillations of the quasiparticle spin, as in Fig. 3 of the main text.
	(a)
	Measured final state probabilities after the application of simultaneous drive pulses, with the spin initialized in either $\oa$ or $\ob$. 
	(b)
	Simulated probabilities. 
	(c) Level diagram with simulation parameters.
	\label{all_rabi}
	}
\end{figure}

Using an experiment similar to that shown in Fig.~\ref{all_raman}, we chose drive frequencies such that the desired Raman process $\oa \leftrightarrow \ob$ was being driven, but the drives were maximally detuned from undesired processes. 
We then varied the amplitudes $A_\downarrow$, $A_\uparrow$ to induce coherent oscillations of the quasiparticle spin [main text, Fig. 3(a)].
To better understand these coherent dynamics, we performed a simulation of our system using QuTiP~\cite{johansson2013qutip} [main text, Fig. 3(b)]. 
Here we present transition probabilities out of the two spin states under the action of these variable amplitude drive pulses, both measured [Fig.~\ref{all_rabi}(a)] and simulated [Fig.~\ref{all_rabi}(b)] (measurements where the system started in $\g$ showed no features). 

The simulation of the coherent dynamics included all four Andreev levels $\oa$, $\ob$, $\oC$, and $\od$ [Fig.~\ref{all_rabi}(c)].
While there were only two drives applied in this experiment, because each drive could couple to each of the four inter-doublet transitions we needed to account for a total of eight Hamiltonian terms. 
Four of these terms produced two Raman processes [dashed double-headed arrows in Fig.~\ref{all_rabi}(c)], one of which was via $\oC$ as desired, and the other was via $\od$. 
The $\oC$ Raman process dominated the dynamics, as the detuning of the drives from $\oC$ was $\Delta_\mathrm{R} = -290~\MHz$, as compared to the detuning to $\od$ which was $\Delta_\mathrm{R}' = 1.36~\GHz$. 
The other four Hamiltonian terms produced Stark shifts [thin solid double-headed arrows in Fig.~\ref{all_rabi}(c)].

The fixed parameters in the simulation were the four inter-doublet transition frequencies, the measured dephasing rates of both doublets, the pulse length and shape (Gaussian, $40~\ns$ standard deviation), and the detunings $\Delta_\mathrm{R} = -290~\MHz$, $\Delta_\mathrm{R}' = 1.36~\GHz$.
We fit the simulation to the data using six free parameters: the four drive matrix elements associated with the four inter-doublet transitions ($M_{\downarrow,\uparrow}$, $M_{\uparrow,\uparrow}$, $M_{\downarrow,\downarrow}$, $M_{\uparrow,\downarrow}$), the detuning $\delta$ from the Raman resonance condition, and the ratio $\alpha$ of the $f_\uparrow$ drive amplitude to the $f_\downarrow$ drive amplitude. 
From the fit, we extract the below values and associated co-variance matrix:

\begin{equation}
    \begin{matrix}
    \delta/(2\pi) = 5.5\\
    \;\;\;M_{\uparrow,\uparrow}/(2\pi) = 232~\MHz \\ 
    \;\;\;M_{\downarrow,\uparrow}/(2\pi) = 255~\MHz \\   \;\;\;M_{\downarrow,\downarrow}/(2\pi) = 280~\MHz \\ 
    \;M_{\uparrow,\downarrow}/(2\pi) = 80~\MHz \\
    \;\;\;\;\;\;\;\;\;\;\alpha = 0.54 \\
    \end{matrix} \;\;\;\;\;\;\;\;
    C = 
    \begin{pmatrix}
      +0.02 & +0.01 & -0.09 & +0.2 &
     +0.2 & +0.0003 \\
      +0.01 & +70 & -50 & -200&
   +200 & +0.02 \\
      -0.09 &-50 & +30 & +100&
  -100 &-0.02 \\
      +0.2 &-200 & +100 & +500&
  -40& -0.05 \\
      +0.2 & +200 & -100 & -40&
   +400 & +0.06 \\
   +0.0003 & +0.02& -0.02 &-0.05&
   +0.06 & +0.00002 \\
    \end{pmatrix}
\end{equation}


\noindent Note that the extracted values of the matrix elements include the drive amplitudes at the device, which we estimate to be $\sim400~\mathrm{nV}$ across the junction.
We note that, in the experimental data, the oscillation minima/maxima do not lie perfectly along $|A_\uparrow| = |A_\downarrow|$; the inclusion of $\delta$ in the fit is necessary to reproduce this. 
The inclusion of the Stark shift terms, on the other hand, were not needed to reproduce the data, but in reality they must be present. 

Additionally, as can be seen in Fig.~\ref{all_rabi}, we observed a drive-induced quasiparticle evaporation rate; the $\g$ population grows as the drive amplitudes are increased. 
In simulation, we found that this quasiparticle evaporation was captured by including two dissipators on the lower doublet of the form $\sqrt{\Gamma_\mathrm{de-trap}}(|A_\downarrow|^2 + |A_\uparrow|^2 + |A_\downarrow||A_\uparrow|) \ket{g}\bra{\downarrow_q}$, $\sqrt{\Gamma_\mathrm{de-trap}}(|A_\downarrow|^2 + |A_\uparrow|^2 + |A_\downarrow||A_\uparrow|) \ket{g}\bra{\uparrow_q}$, with $\Gamma_\mathrm{de-trap}/(2\pi) = 1.2\pm0.1~\MHz$ as extracted from the $\g$ population data.
While this is certainly an over-simplified model (no frequency dependence, no spin dependence, etc.), the scaling of the rate with the drive amplitudes indicates that the evaporation is likely due to multi-photon transitions of the trapped quasiparticle to excited states either in the dot-like levels previously discussed or into the continuum above the superconducting energy gap of the leads. 
This differs with previous results on drive-induced quasiparticle evaporation where a linear scaling of the rate with power was observed~\cite{levenson2014single}, most likely because the Andreev levels studied in this work exist at lower energy. 
These results are also consistent with our observation that the undesired transitions seen in Fig.~\ref{all_rabi} only occur at high powers. 

Finally, we included readout errors in the simulation by extracting the transition probabilities between the outcomes of the first and second readout pulses for the zero-amplitude experimental data and then applying the resultant transfer matrix to the simulated probabilities for all drive amplitudes.
While this did not result in a qualitative change of the data, it was important for replication of the experimental contrast. 
Note that this is why there are some Raman features visible in the simulated $\g$ data: the rate at which the quasiparticle spontaneously evacuates the junction during readout is slightly spin-dependent. 
Although we were unable to measure the upper doublet population directly because the dispersive shift of these states was too small at this bias point, the simulation indicates that it was below 20\%. 

\section{VI. Analysis of the Andreev level coherence times}

\begin{figure}[H]
    \centering
	\includegraphics[scale = 0.075]{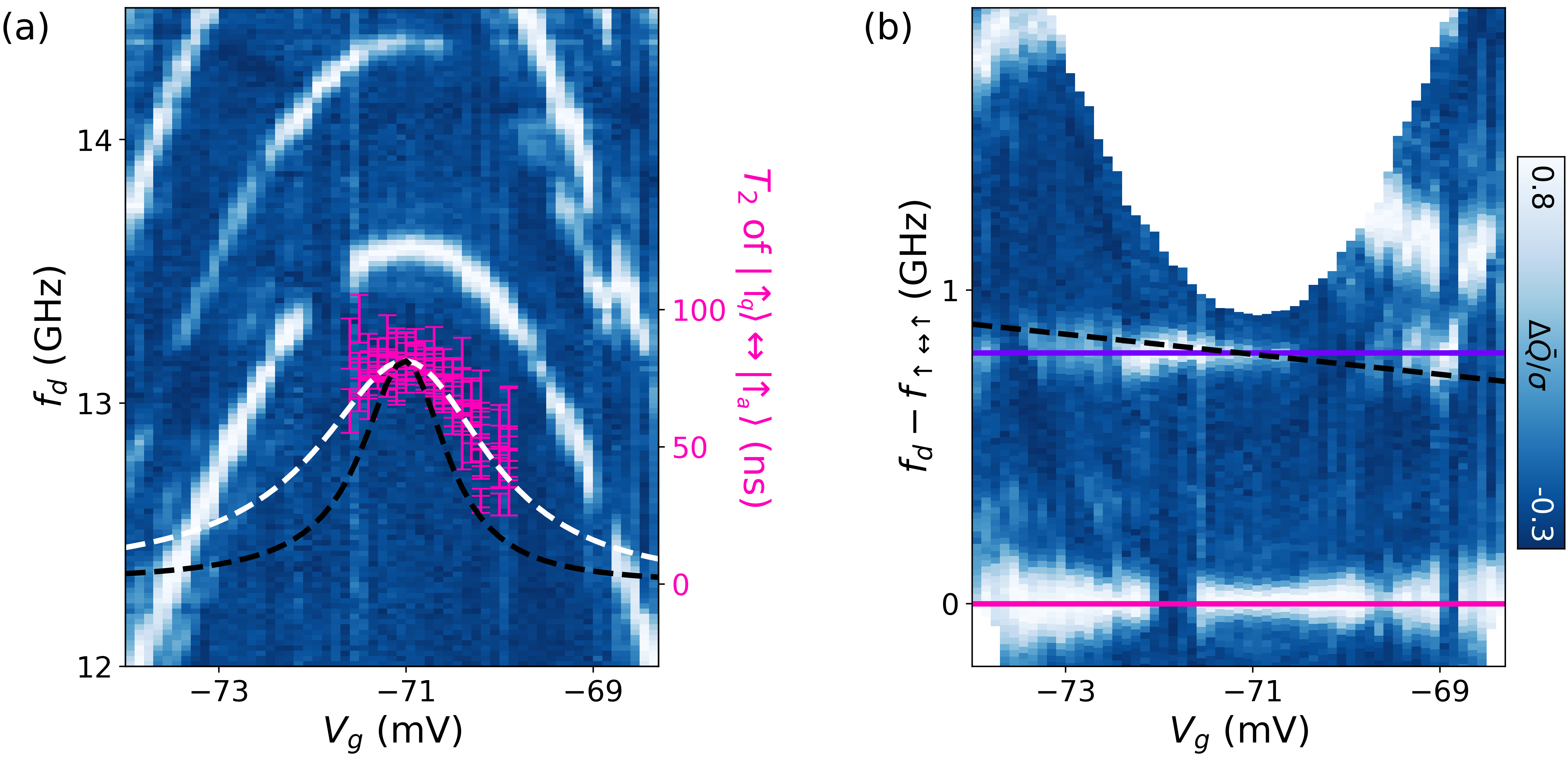} 
	\caption{Extracting a lower-bound  on the electrostatic-noise-induced dephasing time of the quasiparticle spin.  
	(a) 
	Same spectroscopy data as shown in Fig. S2(a). A local maximum is observed in both transitions at $\Vg = -71.0~\mV$. 
	Measurements of the coherence time of $\ob \leftrightarrow \oC$ are shown in pink (right axis).
	White dashed line is a fit to the $T_2$ data assuming first-order noise in $\Vg$ plus a constant dephasing rate. 
	Black dashed line corresponds to the expected $T_2$ given this same constant dephasing rate, but \textit{twice} the $\Vg$ noise. 
	(b)
	Same data as shown in (a), but with the fitted frequency of the $\ob \leftrightarrow \oC$ transition subtracted from $\fd$ for each $\Vg$ bias.
	Both the purple and pink lines have no slope, and lie along the average values of the two transitions. 
	The yellow dashed line is an upper bound on the slope of the $\oa \leftrightarrow \oC$ transition. 
	\label{noise}
	}
\end{figure}

As discussed in the main text, we observed no dependence of the quasiparticle spin $T_2$ on any of the \textit{in-situ} bias knobs ($\Phi,~\Vg,~\Vnw$). 
However, we did find that the coherence times of the inter-doublet transitions depended on $\Vg$ [Fig.~\ref{noise}(a)]. 
In particular, we found that the $T_2$ of the $\ob \leftrightarrow \oC$ was maximum around a $\Vg$ sweet spot at $-71.0~\mV$. 
Away from this sweet spot, first-order electrostatic noise contributed to dephasing, causing $T_2$ to drop.
We model this behavior using the relation for exponential coherence decay $\frac{1}{T_{2}} =  \big(2 \pi V_\mathrm{rms}\frac{df}{d\Vg}\big)^2 + \Gamma_\mathrm{c}$, where $V_\mathrm{rms} = 0.24 \pm 0.01~\mV$ is the effective root-mean-square voltage noise and $\Gamma_\mathrm{c} = 0.012 \pm 0.001~\ns^{-1}$ is a $\Vg$-independent dephasing rate \cite{martinis2009energy}.
We note that the $T_2$ at the sweet spot is given by $\Gamma_\mathrm{c}$, as we found that second-order noise coupling to $\frac{d^2f}{d\Vg^2}$ was negligible \cite{houck2009life}.  

To calculate the effect of this electrostatic noise on the quasiparticle spin $T_2$, we first extracted the $\Vg$ dependence of the $\oa \leftrightarrow \ob$ splitting $\es$, which is given by the difference between the frequencies of the $\oa \leftrightarrow \oC$ and $\ob \leftrightarrow \oC$ transitions. 
We visualize this in Fig.~\ref{noise}(b) by plotting the same two-tone data as shown in Fig.~\ref{noise}(a), but with the fitted value of the $\ob \leftrightarrow \oC$ transition $f_{\uparrow \leftrightarrow \downarrow}$ subtracted from the drive frequency $f_d$ at every $\Vg$ bias. 
The $\ob \leftrightarrow \oC$ transition thus lies along $f_d - f_{\uparrow \leftrightarrow \downarrow} = 0$ (pink horizontal line), while the $\Vg$ dependence of $\es/h$ is given by the behavior of $f_{\downarrow \leftrightarrow \downarrow}$ (purple horizontal line). 
We observe that $\es/h$ has no discernible slope with $\Vg$, consistent with the lack of a spin $T_2$ dependence on $\Vg$. 
However, using an upper bound on this slope $\frac{d\es/h}{d\Vg} < 32~\MHz/\mV$ [black dashed line in Fig.~\ref{noise}(b)] and \textit{twice} the extracted value of $V_\mathrm{rms}$ as an upper bound on the electrostatic noise [black dashed line in Fig.~\ref{noise}(a)], we find a lower bound on the spin dephasing time of $4.2~\us$. 

\section{VII. Analysis of the Andreev level lifetimes}

\begin{figure}[H]
    \centering
	\includegraphics[scale = 0.4]{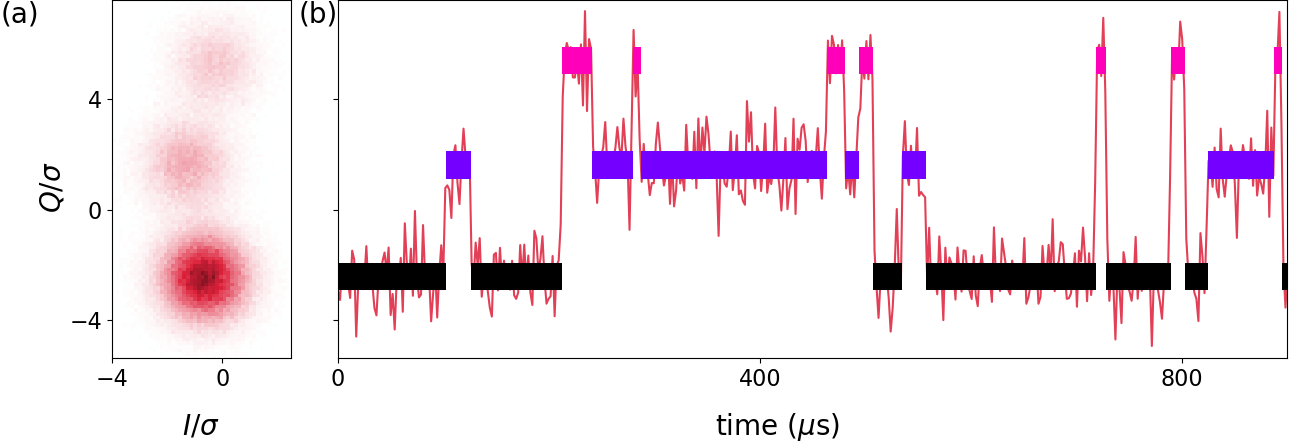} 
	\caption{Quantum jumps of spin and parity at $\Phi = -0.14 \Phi_0$. 
	(a) $\Gamma$ histogram resulting from $10^5$ consecutive readout pulses. 
	(b) Plotting $Q$ against time reveals quantum jumps between $\g$, $\oa$, and $\ob$ (only a subset of the time trace is shown). 
	A hidden Markov algorithm was employed to perform state assignment (colored rectangles), as well as to extract all six transition rates between the three states. 
	These are summarized by the parity lifetime $\Tp = 22\pm1~\us$ and spin lifetime $\Ts = 17\pm1~\us$. 
	\label{quantum_jumps}
	}
\end{figure}

To extract the transition rates between $\g$, $\oa$, and $\ob$, we analyzed the quantum jumps of the system using a hidden Markov model algorithm [Fig.~\ref{quantum_jumps}] \cite{press2015, hays2018direct, hays2020continuous}. 
As we reported in Ref. \cite{hays2020continuous}, the spin lifetime $\Ts$ increased with $|\Phi/\Phi_0|$. 
Unlike in the data presented in Ref. \cite{hays2020continuous}, at this gate bias point the parity lifetime did vary with $\Phi$ due to evacuation of the quasiparticle into the cold, fermionic modes discussed above. 
In Fig.~\ref{quantum_jumps}, we plot quantum jumps for $\Phi = -0.14\Phi_0$ where $\Ts = 17\pm1~\us$ and $\Tp = 22\pm1~\us$. 

\begin{thebibliography}{46}
\expandafter\ifx\csname natexlab\endcsname\relax\def\natexlab#1{#1}\fi
\expandafter\ifx\csname bibnamefont\endcsname\relax
  \def\bibnamefont#1{#1}\fi
\expandafter\ifx\csname bibfnamefont\endcsname\relax
  \def\bibfnamefont#1{#1}\fi
\expandafter\ifx\csname citenamefont\endcsname\relax
  \def\citenamefont#1{#1}\fi
\expandafter\ifx\csname url\endcsname\relax
  \def\url#1{\texttt{#1}}\fi
\expandafter\ifx\csname urlprefix\endcsname\relax\def\urlprefix{URL }\fi
\providecommand{\bibinfo}[2]{#2}
\providecommand{\eprint}[2][]{\url{#2}}

\bibitem[{\citenamefont{Beenakker and Van~Houten}(1991)}]{beenakker1991june}
\bibinfo{author}{\bibfnamefont{C.}~\bibnamefont{Beenakker}} \bibnamefont{and}
  \bibinfo{author}{\bibfnamefont{H.}~\bibnamefont{Van~Houten}},
  \bibinfo{journal}{Phys. Rev. Lett.} \textbf{\bibinfo{volume}{66}},
  \bibinfo{pages}{3056} (\bibinfo{year}{1991}).

\bibitem[{\citenamefont{Furusaki and Tsukada}(1991)}]{Furusaki1991}
\bibinfo{author}{\bibfnamefont{A.}~\bibnamefont{Furusaki}} \bibnamefont{and}
  \bibinfo{author}{\bibfnamefont{M.}~\bibnamefont{Tsukada}},
  \bibinfo{journal}{Phys. Rev. B} \textbf{\bibinfo{volume}{43}},
  \bibinfo{pages}{10164} (\bibinfo{year}{1991}).

\bibitem[{\citenamefont{Clarke and Braginski}(2004)}]{clarke2004}
\bibinfo{editor}{\bibfnamefont{J.}~\bibnamefont{Clarke}} \bibnamefont{and}
  \bibinfo{editor}{\bibfnamefont{A.~I.} \bibnamefont{Braginski}}, eds.,
  \emph{\bibinfo{title}{The {SQUID} {Handbook}}}, vol.~\bibinfo{volume}{1}
  (\bibinfo{publisher}{Wiley, Weinheim}, \bibinfo{year}{2004}).

\bibitem[{\citenamefont{Devoret and Schoelkopf}(2013)}]{devoret2013}
\bibinfo{author}{\bibfnamefont{M.~H.} \bibnamefont{Devoret}} \bibnamefont{and}
  \bibinfo{author}{\bibfnamefont{R.~J.} \bibnamefont{Schoelkopf}},
  \bibinfo{journal}{Science} \textbf{\bibinfo{volume}{339}},
  \bibinfo{pages}{1169} (\bibinfo{year}{2013}).

\bibitem[{\citenamefont{Roy and Devoret}(2016)}]{Roy2016}
\bibinfo{author}{\bibfnamefont{A.}~\bibnamefont{Roy}} \bibnamefont{and}
  \bibinfo{author}{\bibfnamefont{M.}~\bibnamefont{Devoret}},
  \bibinfo{journal}{Comptes Rendus Physique} \textbf{\bibinfo{volume}{17}},
  \bibinfo{pages}{740} (\bibinfo{year}{2016}).

\bibitem[{\citenamefont{Chtchelkatchev and
  Nazarov}(2003)}]{chtchelkatchev2003andreev}
\bibinfo{author}{\bibfnamefont{N.~M.} \bibnamefont{Chtchelkatchev}}
  \bibnamefont{and} \bibinfo{author}{\bibfnamefont{Y.~V.}
  \bibnamefont{Nazarov}}, \bibinfo{journal}{Phys. Rev. Lett.}
  \textbf{\bibinfo{volume}{90}}, \bibinfo{pages}{226806}
  (\bibinfo{year}{2003}).

\bibitem[{\citenamefont{Padurariu and
  Nazarov}(2010)}]{padurariu2010theoretical}
\bibinfo{author}{\bibfnamefont{C.}~\bibnamefont{Padurariu}} \bibnamefont{and}
  \bibinfo{author}{\bibfnamefont{Y.~V.} \bibnamefont{Nazarov}},
  \bibinfo{journal}{Phys. Rev. B} \textbf{\bibinfo{volume}{81}},
  \bibinfo{pages}{144519} (\bibinfo{year}{2010}).

\bibitem[{\citenamefont{Reynoso et~al.}(2012)\citenamefont{Reynoso, Usaj,
  Balseiro, Feinberg, and Avignon}}]{reynoso2012spin}
\bibinfo{author}{\bibfnamefont{A.~A.} \bibnamefont{Reynoso}},
  \bibinfo{author}{\bibfnamefont{G.}~\bibnamefont{Usaj}},
  \bibinfo{author}{\bibfnamefont{C.~A.} \bibnamefont{Balseiro}},
  \bibinfo{author}{\bibfnamefont{D.}~\bibnamefont{Feinberg}}, \bibnamefont{and}
  \bibinfo{author}{\bibfnamefont{M.}~\bibnamefont{Avignon}},
  \bibinfo{journal}{Phys. Rev. B} \textbf{\bibinfo{volume}{86}},
  \bibinfo{pages}{214519} (\bibinfo{year}{2012}).

\bibitem[{\citenamefont{Park and Yeyati}(2017)}]{park2017andreev}
\bibinfo{author}{\bibfnamefont{S.}~\bibnamefont{Park}} \bibnamefont{and}
  \bibinfo{author}{\bibfnamefont{A.~L.} \bibnamefont{Yeyati}},
  \bibinfo{journal}{Phys. Rev. B} \textbf{\bibinfo{volume}{96}},
  \bibinfo{pages}{125416} (\bibinfo{year}{2017}).

\bibitem[{\citenamefont{Hanson et~al.}(2007)\citenamefont{Hanson, Kouwenhoven,
  Petta, Tarucha, and Vandersypen}}]{hanson2007spins}
\bibinfo{author}{\bibfnamefont{R.}~\bibnamefont{Hanson}},
  \bibinfo{author}{\bibfnamefont{L.~P.} \bibnamefont{Kouwenhoven}},
  \bibinfo{author}{\bibfnamefont{J.~R.} \bibnamefont{Petta}},
  \bibinfo{author}{\bibfnamefont{S.}~\bibnamefont{Tarucha}}, \bibnamefont{and}
  \bibinfo{author}{\bibfnamefont{L.~M.~K.} \bibnamefont{Vandersypen}},
  \bibinfo{journal}{Rev. Mod. Phys.} \textbf{\bibinfo{volume}{79}},
  \bibinfo{pages}{1217} (\bibinfo{year}{2007}).

\bibitem[{\citenamefont{Childress and Hanson}(2013)}]{childress2013diamond}
\bibinfo{author}{\bibfnamefont{L.}~\bibnamefont{Childress}} \bibnamefont{and}
  \bibinfo{author}{\bibfnamefont{R.}~\bibnamefont{Hanson}},
  \bibinfo{journal}{MRS Bull.} \textbf{\bibinfo{volume}{38}},
  \bibinfo{pages}{134} (\bibinfo{year}{2013}).

\bibitem[{\citenamefont{Petersson et~al.}(2012)\citenamefont{Petersson, McFaul,
  Schroer, Jung, Taylor, Houck, and Petta}}]{petersson2012circuit}
\bibinfo{author}{\bibfnamefont{K.~D.} \bibnamefont{Petersson}},
  \bibinfo{author}{\bibfnamefont{L.~W.} \bibnamefont{McFaul}},
  \bibinfo{author}{\bibfnamefont{M.~D.} \bibnamefont{Schroer}},
  \bibinfo{author}{\bibfnamefont{M.}~\bibnamefont{Jung}},
  \bibinfo{author}{\bibfnamefont{J.~M.} \bibnamefont{Taylor}},
  \bibinfo{author}{\bibfnamefont{A.~A.} \bibnamefont{Houck}}, \bibnamefont{and}
  \bibinfo{author}{\bibfnamefont{J.~R.} \bibnamefont{Petta}},
  \bibinfo{journal}{Nature} \textbf{\bibinfo{volume}{490}},
  \bibinfo{pages}{380} (\bibinfo{year}{2012}).

\bibitem[{\citenamefont{Samkharadze et~al.}(2018)\citenamefont{Samkharadze,
  Zheng, Kalhor, Brousse, Sammak, Mendes, Blais, Scappucci, and
  Vandersypen}}]{samkharadze2018strong}
\bibinfo{author}{\bibfnamefont{N.}~\bibnamefont{Samkharadze}},
  \bibinfo{author}{\bibfnamefont{G.}~\bibnamefont{Zheng}},
  \bibinfo{author}{\bibfnamefont{N.}~\bibnamefont{Kalhor}},
  \bibinfo{author}{\bibfnamefont{D.}~\bibnamefont{Brousse}},
  \bibinfo{author}{\bibfnamefont{A.}~\bibnamefont{Sammak}},
  \bibinfo{author}{\bibfnamefont{U.}~\bibnamefont{Mendes}},
  \bibinfo{author}{\bibfnamefont{A.}~\bibnamefont{Blais}},
  \bibinfo{author}{\bibfnamefont{G.}~\bibnamefont{Scappucci}},
  \bibnamefont{and}
  \bibinfo{author}{\bibfnamefont{L.}~\bibnamefont{Vandersypen}},
  \bibinfo{journal}{Science} \textbf{\bibinfo{volume}{359}},
  \bibinfo{pages}{1123} (\bibinfo{year}{2018}).

\bibitem[{\citenamefont{Mi et~al.}(2018)\citenamefont{Mi, Benito, Putz, Zajac,
  Taylor, Burkard, and Petta}}]{mi2018coherent}
\bibinfo{author}{\bibfnamefont{X.}~\bibnamefont{Mi}},
  \bibinfo{author}{\bibfnamefont{M.}~\bibnamefont{Benito}},
  \bibinfo{author}{\bibfnamefont{S.}~\bibnamefont{Putz}},
  \bibinfo{author}{\bibfnamefont{D.~M.} \bibnamefont{Zajac}},
  \bibinfo{author}{\bibfnamefont{J.~M.} \bibnamefont{Taylor}},
  \bibinfo{author}{\bibfnamefont{G.}~\bibnamefont{Burkard}}, \bibnamefont{and}
  \bibinfo{author}{\bibfnamefont{J.~R.} \bibnamefont{Petta}},
  \bibinfo{journal}{Nature} \textbf{\bibinfo{volume}{555}},
  \bibinfo{pages}{599} (\bibinfo{year}{2018}).

\bibitem[{\citenamefont{Harvey et~al.}(2018)\citenamefont{Harvey, B\o{}ttcher,
  Orona, Bartlett, Doherty, and Yacoby}}]{harvey2018coupling}
\bibinfo{author}{\bibfnamefont{S.~P.} \bibnamefont{Harvey}},
  \bibinfo{author}{\bibfnamefont{C.~G.~L.} \bibnamefont{B\o{}ttcher}},
  \bibinfo{author}{\bibfnamefont{L.~A.} \bibnamefont{Orona}},
  \bibinfo{author}{\bibfnamefont{S.~D.} \bibnamefont{Bartlett}},
  \bibinfo{author}{\bibfnamefont{A.~C.} \bibnamefont{Doherty}},
  \bibnamefont{and} \bibinfo{author}{\bibfnamefont{A.}~\bibnamefont{Yacoby}},
  \bibinfo{journal}{Phys. Rev. B} \textbf{\bibinfo{volume}{97}},
  \bibinfo{pages}{235409} (\bibinfo{year}{2018}).

\bibitem[{\citenamefont{Landig et~al.}(2018)\citenamefont{Landig, Koski,
  Scarlino, Mendes, Blais, Reichl, Wegscheider, Wallraff, Ensslin, and
  Ihn}}]{landig2018coherent}
\bibinfo{author}{\bibfnamefont{A.~J.} \bibnamefont{Landig}},
  \bibinfo{author}{\bibfnamefont{J.~V.} \bibnamefont{Koski}},
  \bibinfo{author}{\bibfnamefont{P.}~\bibnamefont{Scarlino}},
  \bibinfo{author}{\bibfnamefont{U.}~\bibnamefont{Mendes}},
  \bibinfo{author}{\bibfnamefont{A.}~\bibnamefont{Blais}},
  \bibinfo{author}{\bibfnamefont{C.}~\bibnamefont{Reichl}},
  \bibinfo{author}{\bibfnamefont{W.}~\bibnamefont{Wegscheider}},
  \bibinfo{author}{\bibfnamefont{A.}~\bibnamefont{Wallraff}},
  \bibinfo{author}{\bibfnamefont{K.}~\bibnamefont{Ensslin}}, \bibnamefont{and}
  \bibinfo{author}{\bibfnamefont{T.}~\bibnamefont{Ihn}},
  \bibinfo{journal}{Nature} \textbf{\bibinfo{volume}{560}},
  \bibinfo{pages}{179} (\bibinfo{year}{2018}).

\bibitem[{\citenamefont{Cubaynes et~al.}(2019)\citenamefont{Cubaynes, Delbecq,
  Dartiailh, Assouly, Desjardins, Contamin, Bruhat, Leghtas, Mallet, Cottet
  et~al.}}]{cubaynes2019highly}
\bibinfo{author}{\bibfnamefont{T.}~\bibnamefont{Cubaynes}},
  \bibinfo{author}{\bibfnamefont{M.~R.} \bibnamefont{Delbecq}},
  \bibinfo{author}{\bibfnamefont{M.~C.} \bibnamefont{Dartiailh}},
  \bibinfo{author}{\bibfnamefont{R.}~\bibnamefont{Assouly}},
  \bibinfo{author}{\bibfnamefont{M.~M.} \bibnamefont{Desjardins}},
  \bibinfo{author}{\bibfnamefont{L.~C.} \bibnamefont{Contamin}},
  \bibinfo{author}{\bibfnamefont{L.~E.} \bibnamefont{Bruhat}},
  \bibinfo{author}{\bibfnamefont{Z.}~\bibnamefont{Leghtas}},
  \bibinfo{author}{\bibfnamefont{F.}~\bibnamefont{Mallet}},
  \bibinfo{author}{\bibfnamefont{A.}~\bibnamefont{Cottet}},
  \bibnamefont{et~al.}, \bibinfo{journal}{npj Quantum Inf.}
  \textbf{\bibinfo{volume}{5}}, \bibinfo{pages}{1} (\bibinfo{year}{2019}).

\bibitem[{\citenamefont{Borjans et~al.}(2020)\citenamefont{Borjans, Croot, Mi,
  Gullans, and Petta}}]{borjans2020resonant}
\bibinfo{author}{\bibfnamefont{F.}~\bibnamefont{Borjans}},
  \bibinfo{author}{\bibfnamefont{X.}~\bibnamefont{Croot}},
  \bibinfo{author}{\bibfnamefont{X.}~\bibnamefont{Mi}},
  \bibinfo{author}{\bibfnamefont{M.}~\bibnamefont{Gullans}}, \bibnamefont{and}
  \bibinfo{author}{\bibfnamefont{J.}~\bibnamefont{Petta}},
  \bibinfo{journal}{Nature} \textbf{\bibinfo{volume}{577}},
  \bibinfo{pages}{195} (\bibinfo{year}{2020}).

\bibitem[{\citenamefont{Blais et~al.}(2004)\citenamefont{Blais, Huang,
  Wallraff, Girvin, and Schoelkopf}}]{blais2004cavity}
\bibinfo{author}{\bibfnamefont{A.}~\bibnamefont{Blais}},
  \bibinfo{author}{\bibfnamefont{R.-S.} \bibnamefont{Huang}},
  \bibinfo{author}{\bibfnamefont{A.}~\bibnamefont{Wallraff}},
  \bibinfo{author}{\bibfnamefont{S.~M.} \bibnamefont{Girvin}},
  \bibnamefont{and} \bibinfo{author}{\bibfnamefont{R.~J.}
  \bibnamefont{Schoelkopf}}, \bibinfo{journal}{Phys. Rev. A}
  \textbf{\bibinfo{volume}{69}}, \bibinfo{pages}{062320}
  (\bibinfo{year}{2004}).

\bibitem[{\citenamefont{Wallraff et~al.}(2004)\citenamefont{Wallraff, Schuster,
  Blais, Frunzio, Huang, Majer, Kumar, Girvin, and
  Schoelkopf}}]{wallraff2004strong}
\bibinfo{author}{\bibfnamefont{A.}~\bibnamefont{Wallraff}},
  \bibinfo{author}{\bibfnamefont{D.~I.} \bibnamefont{Schuster}},
  \bibinfo{author}{\bibfnamefont{A.}~\bibnamefont{Blais}},
  \bibinfo{author}{\bibfnamefont{L.}~\bibnamefont{Frunzio}},
  \bibinfo{author}{\bibfnamefont{R.-S.} \bibnamefont{Huang}},
  \bibinfo{author}{\bibfnamefont{J.}~\bibnamefont{Majer}},
  \bibinfo{author}{\bibfnamefont{S.}~\bibnamefont{Kumar}},
  \bibinfo{author}{\bibfnamefont{S.~M.} \bibnamefont{Girvin}},
  \bibnamefont{and} \bibinfo{author}{\bibfnamefont{R.~J.}
  \bibnamefont{Schoelkopf}}, \bibinfo{journal}{Nature}
  \textbf{\bibinfo{volume}{431}}, \bibinfo{pages}{162} (\bibinfo{year}{2004}).

\bibitem[{\citenamefont{Janvier et~al.}(2015)\citenamefont{Janvier, Tosi,
  Bretheau, Girit, Stern, Bertet, Joyez, Vion, Esteve, Goffman
  et~al.}}]{Janvier15}
\bibinfo{author}{\bibfnamefont{C.}~\bibnamefont{Janvier}},
  \bibinfo{author}{\bibfnamefont{L.}~\bibnamefont{Tosi}},
  \bibinfo{author}{\bibfnamefont{L.}~\bibnamefont{Bretheau}},
  \bibinfo{author}{\bibfnamefont{{\c C}.~{\"O}.} \bibnamefont{Girit}},
  \bibinfo{author}{\bibfnamefont{M.}~\bibnamefont{Stern}},
  \bibinfo{author}{\bibfnamefont{P.}~\bibnamefont{Bertet}},
  \bibinfo{author}{\bibfnamefont{P.}~\bibnamefont{Joyez}},
  \bibinfo{author}{\bibfnamefont{D.}~\bibnamefont{Vion}},
  \bibinfo{author}{\bibfnamefont{D.}~\bibnamefont{Esteve}},
  \bibinfo{author}{\bibfnamefont{M.~F.} \bibnamefont{Goffman}},
  \bibnamefont{et~al.}, \bibinfo{journal}{Science}
  \textbf{\bibinfo{volume}{349}}, \bibinfo{pages}{1199} (\bibinfo{year}{2015}).

\bibitem[{\citenamefont{Hays et~al.}(2018)\citenamefont{Hays, de~Lange,
  Serniak, van Woerkom, Bouman, Krogstrup, Nyg{\aa}rd, Geresdi, and
  Devoret}}]{hays2018direct}
\bibinfo{author}{\bibfnamefont{M.}~\bibnamefont{Hays}},
  \bibinfo{author}{\bibfnamefont{G.}~\bibnamefont{de~Lange}},
  \bibinfo{author}{\bibfnamefont{K.}~\bibnamefont{Serniak}},
  \bibinfo{author}{\bibfnamefont{D.}~\bibnamefont{van Woerkom}},
  \bibinfo{author}{\bibfnamefont{D.}~\bibnamefont{Bouman}},
  \bibinfo{author}{\bibfnamefont{P.}~\bibnamefont{Krogstrup}},
  \bibinfo{author}{\bibfnamefont{J.}~\bibnamefont{Nyg{\aa}rd}},
  \bibinfo{author}{\bibfnamefont{A.}~\bibnamefont{Geresdi}}, \bibnamefont{and}
  \bibinfo{author}{\bibfnamefont{M.}~\bibnamefont{Devoret}},
  \bibinfo{journal}{Phys. Rev. Lett.} \textbf{\bibinfo{volume}{121}},
  \bibinfo{pages}{047001} (\bibinfo{year}{2018}).

\bibitem[{\citenamefont{Krogstrup et~al.}(2015)\citenamefont{Krogstrup, Ziino,
  Chang, Albrecht, Madsen, Johnson, Nyg{\aa}rd, Marcus, and
  Jespersen}}]{Krogstrup15}
\bibinfo{author}{\bibfnamefont{P.}~\bibnamefont{Krogstrup}},
  \bibinfo{author}{\bibfnamefont{N.~L.~B.} \bibnamefont{Ziino}},
  \bibinfo{author}{\bibfnamefont{W.}~\bibnamefont{Chang}},
  \bibinfo{author}{\bibfnamefont{S.~M.} \bibnamefont{Albrecht}},
  \bibinfo{author}{\bibfnamefont{M.~H.} \bibnamefont{Madsen}},
  \bibinfo{author}{\bibfnamefont{E.}~\bibnamefont{Johnson}},
  \bibinfo{author}{\bibfnamefont{J.}~\bibnamefont{Nyg{\aa}rd}},
  \bibinfo{author}{\bibfnamefont{C.~M.} \bibnamefont{Marcus}},
  \bibnamefont{and} \bibinfo{author}{\bibfnamefont{T.~S.}
  \bibnamefont{Jespersen}}, \bibinfo{journal}{Nat. Mater.}
  \textbf{\bibinfo{volume}{14}}, \bibinfo{pages}{400} (\bibinfo{year}{2015}).

\bibitem[{\citenamefont{Chang et~al.}(2015)\citenamefont{Chang, Albrecht,
  Jespersen, Kuemmeth, Krogstrup, Nyg{\aa}rd, and Marcus}}]{chang2015hard}
\bibinfo{author}{\bibfnamefont{W.}~\bibnamefont{Chang}},
  \bibinfo{author}{\bibfnamefont{S.}~\bibnamefont{Albrecht}},
  \bibinfo{author}{\bibfnamefont{T.}~\bibnamefont{Jespersen}},
  \bibinfo{author}{\bibfnamefont{F.}~\bibnamefont{Kuemmeth}},
  \bibinfo{author}{\bibfnamefont{P.}~\bibnamefont{Krogstrup}},
  \bibinfo{author}{\bibfnamefont{J.}~\bibnamefont{Nyg{\aa}rd}},
  \bibnamefont{and} \bibinfo{author}{\bibfnamefont{C.~M.}
  \bibnamefont{Marcus}}, \bibinfo{journal}{Nature nanotechnology}
  \textbf{\bibinfo{volume}{10}}, \bibinfo{pages}{232} (\bibinfo{year}{2015}).

\bibitem[{\citenamefont{van Woerkom et~al.}(2017)\citenamefont{van Woerkom,
  Proutski, van Heck, Bouman, V{\"a}yrynen, Glazman, Krogstrup, Nyg{\aa}rd,
  Kouwenhoven, and Geresdi}}]{van2017microwave}
\bibinfo{author}{\bibfnamefont{D.~J.} \bibnamefont{van Woerkom}},
  \bibinfo{author}{\bibfnamefont{A.}~\bibnamefont{Proutski}},
  \bibinfo{author}{\bibfnamefont{B.}~\bibnamefont{van Heck}},
  \bibinfo{author}{\bibfnamefont{D.}~\bibnamefont{Bouman}},
  \bibinfo{author}{\bibfnamefont{J.~I.} \bibnamefont{V{\"a}yrynen}},
  \bibinfo{author}{\bibfnamefont{L.~I.} \bibnamefont{Glazman}},
  \bibinfo{author}{\bibfnamefont{P.}~\bibnamefont{Krogstrup}},
  \bibinfo{author}{\bibfnamefont{J.}~\bibnamefont{Nyg{\aa}rd}},
  \bibinfo{author}{\bibfnamefont{L.~P.} \bibnamefont{Kouwenhoven}},
  \bibnamefont{and} \bibinfo{author}{\bibfnamefont{A.}~\bibnamefont{Geresdi}},
  \bibinfo{journal}{Nat. Phys.} \textbf{\bibinfo{volume}{13}},
  \bibinfo{pages}{876} (\bibinfo{year}{2017}).

\bibitem[{\citenamefont{Tosi et~al.}(2019)\citenamefont{Tosi, Metzger, Goffman,
  Urbina, Pothier, Park, Yeyati, Nyg{\aa}rd, and Krogstrup}}]{tosi2019spin}
\bibinfo{author}{\bibfnamefont{L.}~\bibnamefont{Tosi}},
  \bibinfo{author}{\bibfnamefont{C.}~\bibnamefont{Metzger}},
  \bibinfo{author}{\bibfnamefont{M.}~\bibnamefont{Goffman}},
  \bibinfo{author}{\bibfnamefont{C.}~\bibnamefont{Urbina}},
  \bibinfo{author}{\bibfnamefont{H.}~\bibnamefont{Pothier}},
  \bibinfo{author}{\bibfnamefont{S.}~\bibnamefont{Park}},
  \bibinfo{author}{\bibfnamefont{A.~L.} \bibnamefont{Yeyati}},
  \bibinfo{author}{\bibfnamefont{J.}~\bibnamefont{Nyg{\aa}rd}},
  \bibnamefont{and}
  \bibinfo{author}{\bibfnamefont{P.}~\bibnamefont{Krogstrup}},
  \bibinfo{journal}{Phys. Rev. X} \textbf{\bibinfo{volume}{9}},
  \bibinfo{pages}{011010} (\bibinfo{year}{2019}).

\bibitem[{\citenamefont{Larsen et~al.}(2015)\citenamefont{Larsen, Petersson,
  Kuemmeth, Jespersen, Krogstrup, Nyg{\aa}rd, and
  Marcus}}]{larsen2015semiconductor}
\bibinfo{author}{\bibfnamefont{T.~W.} \bibnamefont{Larsen}},
  \bibinfo{author}{\bibfnamefont{K.~D.} \bibnamefont{Petersson}},
  \bibinfo{author}{\bibfnamefont{F.}~\bibnamefont{Kuemmeth}},
  \bibinfo{author}{\bibfnamefont{T.~S.} \bibnamefont{Jespersen}},
  \bibinfo{author}{\bibfnamefont{P.}~\bibnamefont{Krogstrup}},
  \bibinfo{author}{\bibfnamefont{J.}~\bibnamefont{Nyg{\aa}rd}},
  \bibnamefont{and} \bibinfo{author}{\bibfnamefont{C.~M.}
  \bibnamefont{Marcus}}, \bibinfo{journal}{Phys. Rev. Lett.}
  \textbf{\bibinfo{volume}{115}}, \bibinfo{pages}{127001}
  (\bibinfo{year}{2015}).

\bibitem[{\citenamefont{De~Lange et~al.}(2015)\citenamefont{De~Lange, Van~Heck,
  Bruno, Van~Woerkom, Geresdi, Plissard, Bakkers, Akhmerov, and
  DiCarlo}}]{de2015realization}
\bibinfo{author}{\bibfnamefont{G.}~\bibnamefont{De~Lange}},
  \bibinfo{author}{\bibfnamefont{B.}~\bibnamefont{Van~Heck}},
  \bibinfo{author}{\bibfnamefont{A.}~\bibnamefont{Bruno}},
  \bibinfo{author}{\bibfnamefont{D.}~\bibnamefont{Van~Woerkom}},
  \bibinfo{author}{\bibfnamefont{A.}~\bibnamefont{Geresdi}},
  \bibinfo{author}{\bibfnamefont{S.}~\bibnamefont{Plissard}},
  \bibinfo{author}{\bibfnamefont{E.}~\bibnamefont{Bakkers}},
  \bibinfo{author}{\bibfnamefont{A.}~\bibnamefont{Akhmerov}}, \bibnamefont{and}
  \bibinfo{author}{\bibfnamefont{L.}~\bibnamefont{DiCarlo}},
  \bibinfo{journal}{Phys. Rev. Lett.} \textbf{\bibinfo{volume}{115}},
  \bibinfo{pages}{127002} (\bibinfo{year}{2015}).

\bibitem[{\citenamefont{Fu and Kane}(2008)}]{fu2008superconducting}
\bibinfo{author}{\bibfnamefont{L.}~\bibnamefont{Fu}} \bibnamefont{and}
  \bibinfo{author}{\bibfnamefont{C.~L.} \bibnamefont{Kane}},
  \bibinfo{journal}{Phys. Rev. Lett.} \textbf{\bibinfo{volume}{100}},
  \bibinfo{pages}{096407} (\bibinfo{year}{2008}).

\bibitem[{\citenamefont{Lutchyn et~al.}(2010)\citenamefont{Lutchyn, Sau, and
  Das~Sarma}}]{Lutchyn2010}
\bibinfo{author}{\bibfnamefont{R.~M.} \bibnamefont{Lutchyn}},
  \bibinfo{author}{\bibfnamefont{J.~D.} \bibnamefont{Sau}}, \bibnamefont{and}
  \bibinfo{author}{\bibfnamefont{S.}~\bibnamefont{Das~Sarma}},
  \bibinfo{journal}{Phys. Rev. Lett.} \textbf{\bibinfo{volume}{105}},
  \bibinfo{pages}{077001} (\bibinfo{year}{2010}).

\bibitem[{\citenamefont{Oreg et~al.}(2010)\citenamefont{Oreg, Refael, and von
  Oppen}}]{Oreg2010}
\bibinfo{author}{\bibfnamefont{Y.}~\bibnamefont{Oreg}},
  \bibinfo{author}{\bibfnamefont{G.}~\bibnamefont{Refael}}, \bibnamefont{and}
  \bibinfo{author}{\bibfnamefont{F.}~\bibnamefont{von Oppen}},
  \bibinfo{journal}{Phys. Rev. Lett.} \textbf{\bibinfo{volume}{105}},
  \bibinfo{pages}{177002} (\bibinfo{year}{2010}).

\bibitem[{\citenamefont{Mourik et~al.}(2012)\citenamefont{Mourik, Zuo, Frolov,
  Plissard, Bakkers, and Kouwenhoven}}]{Mourik2012}
\bibinfo{author}{\bibfnamefont{V.}~\bibnamefont{Mourik}},
  \bibinfo{author}{\bibfnamefont{K.}~\bibnamefont{Zuo}},
  \bibinfo{author}{\bibfnamefont{S.~M.} \bibnamefont{Frolov}},
  \bibinfo{author}{\bibfnamefont{S.~R.} \bibnamefont{Plissard}},
  \bibinfo{author}{\bibfnamefont{E.~P. A.~M.} \bibnamefont{Bakkers}},
  \bibnamefont{and} \bibinfo{author}{\bibfnamefont{L.~P.}
  \bibnamefont{Kouwenhoven}}, \bibinfo{journal}{Science}
  \textbf{\bibinfo{volume}{336}}, \bibinfo{pages}{1003} (\bibinfo{year}{2012}).

\bibitem[{\citenamefont{Deng et~al.}(2016)\citenamefont{Deng, Vaitiek{\.e}nas,
  Hansen, Danon, Leijnse, Flensberg, Nyg{\aa}rd, Krogstrup, and
  Marcus}}]{deng2016majorana}
\bibinfo{author}{\bibfnamefont{M.}~\bibnamefont{Deng}},
  \bibinfo{author}{\bibfnamefont{S.}~\bibnamefont{Vaitiek{\.e}nas}},
  \bibinfo{author}{\bibfnamefont{E.~B.} \bibnamefont{Hansen}},
  \bibinfo{author}{\bibfnamefont{J.}~\bibnamefont{Danon}},
  \bibinfo{author}{\bibfnamefont{M.}~\bibnamefont{Leijnse}},
  \bibinfo{author}{\bibfnamefont{K.}~\bibnamefont{Flensberg}},
  \bibinfo{author}{\bibfnamefont{J.}~\bibnamefont{Nyg{\aa}rd}},
  \bibinfo{author}{\bibfnamefont{P.}~\bibnamefont{Krogstrup}},
  \bibnamefont{and} \bibinfo{author}{\bibfnamefont{C.~M.}
  \bibnamefont{Marcus}}, \bibinfo{journal}{Science}
  \textbf{\bibinfo{volume}{354}}, \bibinfo{pages}{1557} (\bibinfo{year}{2016}).

\bibitem[{\citenamefont{Hays et~al.}(2020)\citenamefont{Hays, Fatemi, Serniak,
  Bouman, Diamond, de~Lange, Krogstrup, Nyg{\aa}rd, Geresdi, and
  Devoret}}]{hays2020continuous}
\bibinfo{author}{\bibfnamefont{M.}~\bibnamefont{Hays}},
  \bibinfo{author}{\bibfnamefont{V.}~\bibnamefont{Fatemi}},
  \bibinfo{author}{\bibfnamefont{K.}~\bibnamefont{Serniak}},
  \bibinfo{author}{\bibfnamefont{D.}~\bibnamefont{Bouman}},
  \bibinfo{author}{\bibfnamefont{S.}~\bibnamefont{Diamond}},
  \bibinfo{author}{\bibfnamefont{G.}~\bibnamefont{de~Lange}},
  \bibinfo{author}{\bibfnamefont{P.}~\bibnamefont{Krogstrup}},
  \bibinfo{author}{\bibfnamefont{J.}~\bibnamefont{Nyg{\aa}rd}},
  \bibinfo{author}{\bibfnamefont{A.}~\bibnamefont{Geresdi}}, \bibnamefont{and}
  \bibinfo{author}{\bibfnamefont{M.}~\bibnamefont{Devoret}},
  \bibinfo{journal}{Nat. Phys.} \textbf{\bibinfo{volume}{16}},
  \bibinfo{pages}{1103} (\bibinfo{year}{2020}).

\bibitem[{\citenamefont{Nadj-Perge et~al.}(2010)\citenamefont{Nadj-Perge,
  Frolov, Bakkers, and Kouwenhoven}}]{nadj2010spin}
\bibinfo{author}{\bibfnamefont{S.}~\bibnamefont{Nadj-Perge}},
  \bibinfo{author}{\bibfnamefont{S.}~\bibnamefont{Frolov}},
  \bibinfo{author}{\bibfnamefont{E.}~\bibnamefont{Bakkers}}, \bibnamefont{and}
  \bibinfo{author}{\bibfnamefont{L.~P.} \bibnamefont{Kouwenhoven}},
  \bibinfo{journal}{Nature} \textbf{\bibinfo{volume}{468}},
  \bibinfo{pages}{1084} (\bibinfo{year}{2010}).

\bibitem[{\citenamefont{Johansson et~al.}(2013)\citenamefont{Johansson, Nation,
  and Nori}}]{johansson2013qutip}
\bibinfo{author}{\bibfnamefont{J.~R.} \bibnamefont{Johansson}},
  \bibinfo{author}{\bibfnamefont{P.~D.} \bibnamefont{Nation}},
  \bibnamefont{and} \bibinfo{author}{\bibfnamefont{F.}~\bibnamefont{Nori}},
  \bibinfo{journal}{Computer Physics Communications}
  \textbf{\bibinfo{volume}{184}}, \bibinfo{pages}{1234} (\bibinfo{year}{2013}).

\bibitem[{\citenamefont{Van~den Berg et~al.}(2013)\citenamefont{Van~den Berg,
  Nadj-Perge, Pribiag, Plissard, Bakkers, Frolov, and
  Kouwenhoven}}]{van2013fast}
\bibinfo{author}{\bibfnamefont{J.}~\bibnamefont{Van~den Berg}},
  \bibinfo{author}{\bibfnamefont{S.}~\bibnamefont{Nadj-Perge}},
  \bibinfo{author}{\bibfnamefont{V.}~\bibnamefont{Pribiag}},
  \bibinfo{author}{\bibfnamefont{S.}~\bibnamefont{Plissard}},
  \bibinfo{author}{\bibfnamefont{E.}~\bibnamefont{Bakkers}},
  \bibinfo{author}{\bibfnamefont{S.}~\bibnamefont{Frolov}}, \bibnamefont{and}
  \bibinfo{author}{\bibfnamefont{L.}~\bibnamefont{Kouwenhoven}},
  \bibinfo{journal}{Phys. Rev. Lett.} \textbf{\bibinfo{volume}{110}},
  \bibinfo{pages}{066806} (\bibinfo{year}{2013}).

\bibitem[{\citenamefont{Bretheau et~al.}(2013)\citenamefont{Bretheau, Girit,
  Pothier, Esteve, and Urbina}}]{bretheau2013exciting}
\bibinfo{author}{\bibfnamefont{L.}~\bibnamefont{Bretheau}},
  \bibinfo{author}{\bibfnamefont{{\c{C}}.}~\bibnamefont{Girit}},
  \bibinfo{author}{\bibfnamefont{H.}~\bibnamefont{Pothier}},
  \bibinfo{author}{\bibfnamefont{D.}~\bibnamefont{Esteve}}, \bibnamefont{and}
  \bibinfo{author}{\bibfnamefont{C.}~\bibnamefont{Urbina}},
  \bibinfo{journal}{Nature} \textbf{\bibinfo{volume}{499}},
  \bibinfo{pages}{312} (\bibinfo{year}{2013}).

\bibitem[{\citenamefont{Malinowski et~al.}(2017)\citenamefont{Malinowski,
  Martins, Nissen, Barnes, Cywi{\'n}ski, Rudner, Fallahi, Gardner, Manfra,
  Marcus et~al.}}]{malinowski2017notch}
\bibinfo{author}{\bibfnamefont{F.~K.} \bibnamefont{Malinowski}},
  \bibinfo{author}{\bibfnamefont{F.}~\bibnamefont{Martins}},
  \bibinfo{author}{\bibfnamefont{P.~D.} \bibnamefont{Nissen}},
  \bibinfo{author}{\bibfnamefont{E.}~\bibnamefont{Barnes}},
  \bibinfo{author}{\bibfnamefont{{\L}.}~\bibnamefont{Cywi{\'n}ski}},
  \bibinfo{author}{\bibfnamefont{M.~S.} \bibnamefont{Rudner}},
  \bibinfo{author}{\bibfnamefont{S.}~\bibnamefont{Fallahi}},
  \bibinfo{author}{\bibfnamefont{G.~C.} \bibnamefont{Gardner}},
  \bibinfo{author}{\bibfnamefont{M.~J.} \bibnamefont{Manfra}},
  \bibinfo{author}{\bibfnamefont{C.~M.} \bibnamefont{Marcus}},
  \bibnamefont{et~al.}, \bibinfo{journal}{Nature nanotechnology}
  \textbf{\bibinfo{volume}{12}}, \bibinfo{pages}{16} (\bibinfo{year}{2017}).

\bibitem[{\citenamefont{Cerrillo et~al.}(2020)\citenamefont{Cerrillo, Hays,
  Fatemi, and Levy~Yeyati}}]{cerrillo2020}
\bibinfo{author}{\bibfnamefont{J.}~\bibnamefont{Cerrillo}},
  \bibinfo{author}{\bibfnamefont{M.}~\bibnamefont{Hays}},
  \bibinfo{author}{\bibfnamefont{V.}~\bibnamefont{Fatemi}}, \bibnamefont{and}
  \bibinfo{author}{\bibfnamefont{A.}~\bibnamefont{Levy~Yeyati}},
  \bibinfo{journal}{arXiv preprint: 2012.07132}  (\bibinfo{year}{2020}).

\bibitem[{\citenamefont{Serniak et~al.}(2019)\citenamefont{Serniak, Diamond,
  Hays, Fatemi, Shankar, Frunzio, Schoelkopf, and Devoret}}]{serniak2019direct}
\bibinfo{author}{\bibfnamefont{K.}~\bibnamefont{Serniak}},
  \bibinfo{author}{\bibfnamefont{S.}~\bibnamefont{Diamond}},
  \bibinfo{author}{\bibfnamefont{M.}~\bibnamefont{Hays}},
  \bibinfo{author}{\bibfnamefont{V.}~\bibnamefont{Fatemi}},
  \bibinfo{author}{\bibfnamefont{S.}~\bibnamefont{Shankar}},
  \bibinfo{author}{\bibfnamefont{L.}~\bibnamefont{Frunzio}},
  \bibinfo{author}{\bibfnamefont{R.}~\bibnamefont{Schoelkopf}},
  \bibnamefont{and} \bibinfo{author}{\bibfnamefont{M.}~\bibnamefont{Devoret}},
  \bibinfo{journal}{Phys. Rev. Applied} \textbf{\bibinfo{volume}{12}},
  \bibinfo{pages}{014052} (\bibinfo{year}{2019}).
  
\bibitem{frattini2018}
\bibinfo{author}{Frattini, N.~E.}, \bibinfo{author}{Sivak, V.~V.},
  \bibinfo{author}{Lingenfelter, A.}, \bibinfo{author}{Shankar, S.} \&
  \bibinfo{author}{Devoret, M.~H.},
\bibinfo{journal}{Phys. Rev. Applied}
  \textbf{\bibinfo{volume}{10}}, \bibinfo{pages}{054020}
  (\bibinfo{year}{2018}).

\bibitem[{\citenamefont{Governale and Z{\"u}licke}(2002)}]{governale2002spin}
\bibinfo{author}{\bibfnamefont{M.}~\bibnamefont{Governale}} \bibnamefont{and}
  \bibinfo{author}{\bibfnamefont{U.}~\bibnamefont{Z{\"u}licke}},
  \bibinfo{journal}{Phys. Rev. B} \textbf{\bibinfo{volume}{66}},
  \bibinfo{pages}{073311} (\bibinfo{year}{2002}).

\bibitem[{\citenamefont{Levenson-Falk et~al.}(2014)\citenamefont{Levenson-Falk,
  Kos, Vijay, Glazman, and Siddiqi}}]{levenson2014single}
\bibinfo{author}{\bibfnamefont{E.}~\bibnamefont{Levenson-Falk}},
  \bibinfo{author}{\bibfnamefont{F.}~\bibnamefont{Kos}},
  \bibinfo{author}{\bibfnamefont{R.}~\bibnamefont{Vijay}},
  \bibinfo{author}{\bibfnamefont{L.}~\bibnamefont{Glazman}}, \bibnamefont{and}
  \bibinfo{author}{\bibfnamefont{I.}~\bibnamefont{Siddiqi}},
  \bibinfo{journal}{Phys. Rev. Lett.} \textbf{\bibinfo{volume}{112}},
  \bibinfo{pages}{047002} (\bibinfo{year}{2014}).

\bibitem[{\citenamefont{Martinis et~al.}(2009)\citenamefont{Martinis, Ansmann,
  and Aumentado}}]{martinis2009energy}
\bibinfo{author}{\bibfnamefont{J.~M.} \bibnamefont{Martinis}},
  \bibinfo{author}{\bibfnamefont{S.}~\bibnamefont{Nam}}, 
\bibinfo{author}{\bibfnamefont{J.}~\bibnamefont{Aumentado}}, 
  \bibinfo{author}{\bibfnamefont{K.~M.} \bibnamefont{Lang}},\bibnamefont{ and}
  \bibinfo{author}{\bibfnamefont{C.}~\bibnamefont{Urbina}},
  \bibinfo{journal}{Phys. Rev. B} \textbf{\bibinfo{volume}{67}},
  \bibinfo{pages}{094510} (\bibinfo{year}{2003}).

\bibitem[{\citenamefont{Houck et~al.}(2009)\citenamefont{Houck, Koch, Devoret,
  Girvin, and Schoelkopf}}]{houck2009life}
\bibinfo{author}{\bibfnamefont{A.~A.} \bibnamefont{Houck}},
  \bibinfo{author}{\bibfnamefont{J.}~\bibnamefont{Koch}},
  \bibinfo{author}{\bibfnamefont{M.~H.} \bibnamefont{Devoret}},
  \bibinfo{author}{\bibfnamefont{S.~M.} \bibnamefont{Girvin}},
  \bibnamefont{and} \bibinfo{author}{\bibfnamefont{R.~J.}
  \bibnamefont{Schoelkopf}}, \bibinfo{journal}{Quantum Information Processing}
  \textbf{\bibinfo{volume}{8}}, \bibinfo{pages}{105} (\bibinfo{year}{2009}).

\bibitem[{\citenamefont{Press}(2002)}]{press2015}
\bibinfo{author}{\bibfnamefont{W.}~\bibnamefont{Press}},
  \emph{\bibinfo{title}{Numerical Recipes in C++ : The Art of Scientific
  Computing}} (\bibinfo{publisher}{Cambridge University Press},
  \bibinfo{year}{2002}).
  


\end{thebibliography}
\end{document}